%% file: main.tex
\documentclass[runningheads]{llncs}
\usepackage{amsmath}
\usepackage{cite}
\usepackage{calrsfs}
\usepackage{graphicx}
\usepackage{tabularx}
\usepackage{xcolor}
\usepackage{color,soul}
\usepackage{subcaption}
\usepackage{booktabs}
\usepackage[english]{babel}
\usepackage{float}
\usepackage{framed}
\usepackage{algorithmicx}
\usepackage{algpseudocode}
   \usepackage{algorithmicx}
    \usepackage[ruled]{algorithm}
    \usepackage{algpseudocode}
\alglanguage{pseudocode}

\usepackage{hyperref}
\begin{document}
%
%\title{Interference-aware Inference pipelines}%\thanks{Supported by organization x.}
%
\title{ODIN: Overcoming Dynamic Interference in iNference pipelines}
\titlerunning{ODIN: Overcoming Dynamic Interference in iNference pipelines}

\author{Pirah Noor Soomro\orcidID{0000-0001-8654-3249} \and Nikela Papadopoulou\orcidID{0000-0003-2141-5654} \and Miquel Pericàs\orcidID{0000-0002-7583-6609}}
\institute{Chalmers University of Technology, 
Gothenburg, Sweden \\ 
\email{\{pirah,nikela,miquelp\}@chalmers.se}}
\authorrunning{P. Noor Soomro et al.}
\maketitle              % typeset the header of the contribution

\begin{abstract}
As an increasing number of businesses becomes powered by machine-learning, inference becomes a core operation, with a growing trend to be offered as a service. In this context, the inference task must meet certain service-level objectives (SLOs), such as high throughput and low latency. However, these targets can be compromised by interference caused by long- or short-lived co-located tasks. Prior works focus on the generic problem of co-scheduling to mitigate the effect of interference on the performance-critical task. In this work, we focus on inference pipelines and propose ODIN, a technique to mitigate the effect of interference on the performance of the inference task, based on the online scheduling of the pipeline stages. Our technique detects interference online and automatically re-balances the pipeline stages to mitigate the performance degradation of the inference task. We demonstrate that ODIN successfully mitigates the effect of interference, sustaining the latency and throughput of CNN inference, and outperforms the least-loaded scheduling (LLS), a common technique for interference mitigation. Additionally, it is effective in maintaining service-level objectives for inference, and it is scalable to large network models executing on multiple processing elements.

 \end{abstract}

\input{shisha/introduction}
\input{shisha/motivation}
\input{shisha/background}
%\input{shisha/relatedWork}
\input{shisha/methodology}
%\input{shisha/experimental_setup}
\input{shisha/evaluation}

\input{shisha/conclusion}
\bibliographystyle{splncs04}

\bibliography{references}
%
%
%


\end{document}

%% file: shisha/introduction.tex
\section{Introduction}

As machine learning becomes the backbone of the digital world, there is an increasing demand for predictions as a service. This has led to the advent of inference-serving systems~\cite{crankshaw2017clipper, lee2018pretzel, olston2017tensorflow, liberty2020elastic, romero2021infaas}. These systems deploy pre-trained model pipelines, i.e. inference pipelines, on the cloud, serving inference queries to users and applications, often under strict quality-of-service (QoS) requirements for the response times and throughput of the queries\cite{zhang2019mark}, expressed as service level objectives (SLOs).  However, due to the limited availability of resources of cloud systems, in combination with high demand, inference pipelines are often co-located with other workloads, either as part of the inference-serving system, which may opt to co-locate multiple inference pipelines \cite{yeung2020horus, mendoza2021interference}, or as part of common multi-tenancy practices of cloud providers \cite{delimitrou2013paragon, delimitrou2014quasar} to increase utilization. The resulting interference from the co-located workload can have devastating effects on inference performance, leading to violation of the SLOs. \looseness=-1

The mitigation of the effect of interference from co-located workloads on the performance of a critical application has been studied extensively. Several scheduling techniques focus on the generic problem of workload colocation, trying to retain or guarantee the performance of one critical or high-priority workload under interference~\cite{delimitrou2013paragon, delimitrou2014quasar, chen2016baymax, chen2017prophet}, while more recent works focus on the problem of colocating inference pipelines specifically \cite{romero2021infaas, mendoza2021interference, ke2022hercules}. Most of these techniques perform extensive offline profiling and/or characterization of workloads and workload colocations, and build pre-trained machine-learning models or analytical models for each system, while a brief profiling phase may also be required to characterize a workload \cite{delimitrou2013paragon,delimitrou2014quasar}. These techniques proactively partition resources to the workloads to mitigate the effect of interference, but may reactively repartition resources or evict colocated workloads in response to changes in the observed performance or interference. Finally, some techniques only focus on interference effects affecting specific resources, such as GPU accelerators~\cite{chen2016baymax,chen2017prophet}. \looseness=-1

%In interference-free environments, a 
One way to achieve high throughput and low latency for inference pipelines is pipeline parallelism. Pipeline parallelism in the form of layer pipelining has been used extensively in training~\cite{narayanan2019pipedream, huang2019gpipe, fan2021dapple, li2021chimera}, and in inference~\cite{wang2019high, kang2020scheduling}, in combination with operator parallelism, as it is able to reduce data movement costs. To exploit pipelined parallelism, several techniques focus on finding near-optimal pipeline schedules online, using heuristics to tackle the large search space \cite{jeong2021deep, soomro2021online, soomro2022shisha, chang2022pipebert}. The ability to rebalance pipeline stages online leaves ample room for the optimization of the execution of a pipeline under the presence of interference, where such a reactive technique can detect and mitigate performance degradation, by making better utilization of the existing resources. 

In this work, we propose ODIN, an online solution that dynamically detects interference and adapts the execution of inference pipelines on a given set of processing elements. Thus, inference-serving systems can exploit them to reduce SLO violations in the presence of interference without eviction or resource repartitioning. ODIN does not require offline resource utilization profiles for the inference, and relies only on runtime observed execution times of pipeline stages, therefore being easily applicable to any system. Additionally, ODIN avoids the costly process of building system-specific or pipeline-specific models to characterize interference. Instead, it dynamically reacts and adapts to the presence of interference while executing the inference pipeline. ODIN by itself does not have a notion of SLOs. It is a best-effort solution to quickly achieve near-optimal throughput and latency in the presence of interference, which thereby results in improved SLO conformance compared to a baseline least-loaded scheduler (LLS). \looseness=-1
%Given enough resources this allows the inference system to regain SLO conformance.
%, it makes optimal use of the system resources, even when they are constrained by a colocated application causing interference, in a best-effort approach to retain the throughput and latency of the inference pipeline. 

ODIN employs a heuristic pipeline scheduling algorithm, which uses the execution times of pipeline stages, compares them against interference-free performance values, and then moves network layers between pipeline stages, with the goal to reduce the work on the execution unit affected by interference, while maximizing the overall throughput of the pipeline. To minimize the duration of the mitigation phase and quickly react to performance changes due to interference, the heuristic takes into account the extent of the performance degradation. We extensively test ODIN with 12 different scenarios of interference in 9 different frequency-duration settings and compare against the baseline least-loaded scheduler (LLS), which selects the least-loaded execution unit to assign work to. Our experiments show that ODIN sustains high throughput and low latency, including tail latency, under the different interference scenarios, and reacts quickly with a short mitigation phase, which takes 5-15 timesteps, outperforming LLS by 15\% in latency and 20\% in throughput on average. 
Additionally, with an SLO set at 80\% of the original throughput, our solution is able to avoid 80\% of SLO violations under interference, in contrast to LLS, which only delivers 50\% SLO conformance. We also test the scalability of ODIN with a deep neural network model on highly parallel platforms, showing that the quality of the solution is independent of the number of execution units and depth of neural network. \looseness=-1

%% file: shisha/motivation.tex
\section{Background and Motivation}
%1: about inference pipelines
Parallel inference pipelines provide a way to maximize the throughput of inference applications, as layer-wise parallelism offers reduced communication and minimizes the need to copy weights between execution units \cite{ben2019demystifying}. The parallelism exposed in parallel inference pipelines is across layers, with each layer being assigned to a pipeline stage, as well as within layers, where operators are parallelized for faster execution.  
%Therefore it is one of the common ways to run commercial inference-based applications such as recommendation systems[references]. 
%Parallel pipelines cover a broad spectrum of parallelism, such as parallelism within and across the layers of the network. Hence there is a potential to adjust the assignment of workload to execution units in order to maximize the throughput of the pipeline. 
%2: about type of intereference we are looking at
A common way to execute pipelines is the ``bind-to-stage" approach \cite{lee2015fly}, where each stage of the pipeline is assigned to a unique set of compute units, i.e. an execution place, without sharing resources with other stages. In our work, we also assume that execution places do not share resources, therefore a pipeline stage will not experience interference from pipeline stages running on other execution places. To achieve high throughput, the pipeline stages need to be balanced, otherwise, throughput becomes limited by pipeline stalls, as the pipeline stages have a linear dependence.

\begin{figure}[!tb]
\centering
\begin{minipage}[t]{.22\textwidth}
    \centering
    \begin{subfigure}[t]{\textwidth}
        \includegraphics[width=0.99\linewidth]{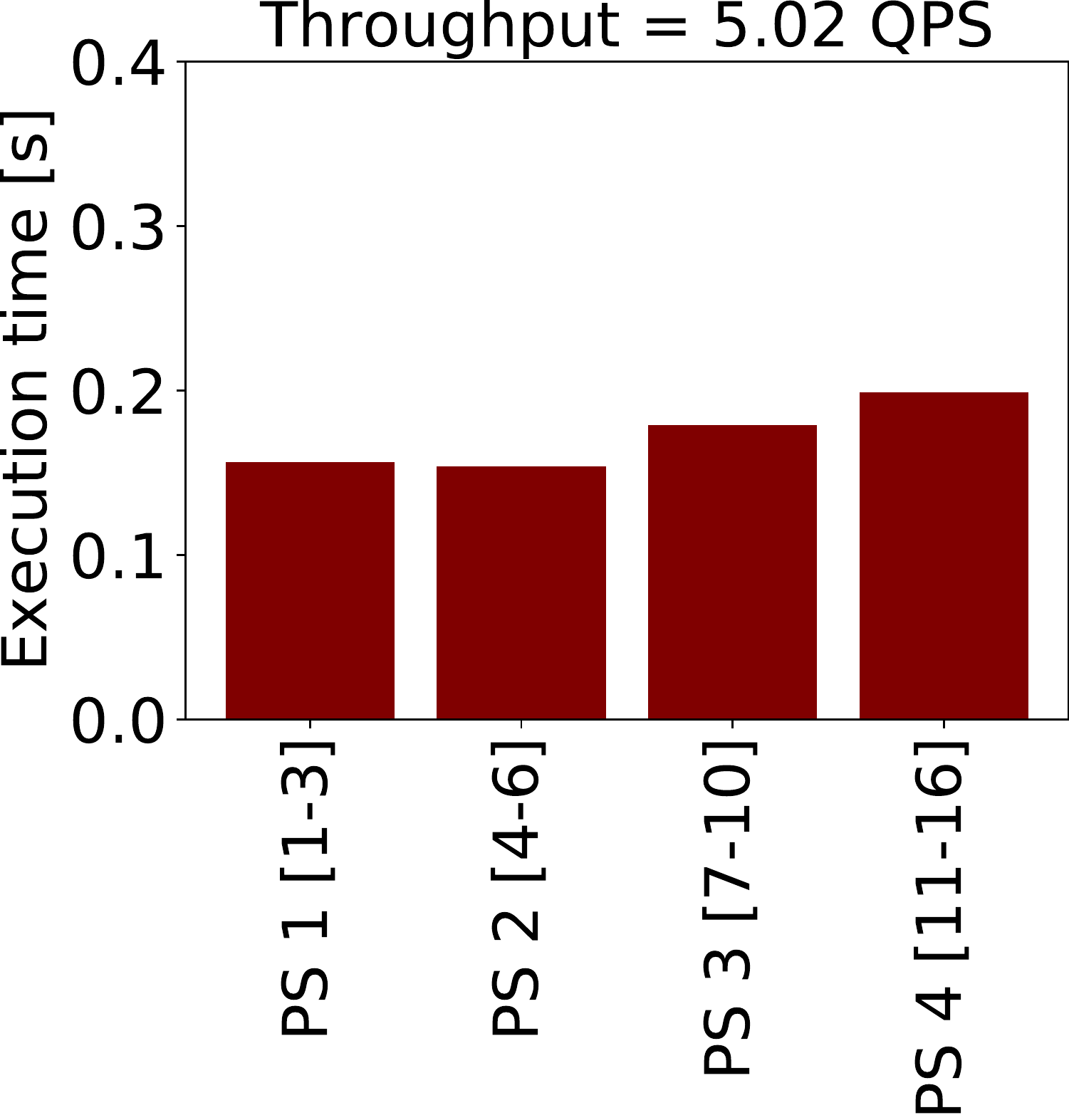}
        \caption{No interference}
        %Execution on 4 execution places without interference.
        \label{fig:motivation_2_case_1}
    \end{subfigure}
\end{minipage}
\hspace{0.3pt}
\begin{minipage}[t]{.22\linewidth}
  \centering
  \begin{subfigure}[t]{\textwidth}
        \includegraphics[width=0.99\linewidth]{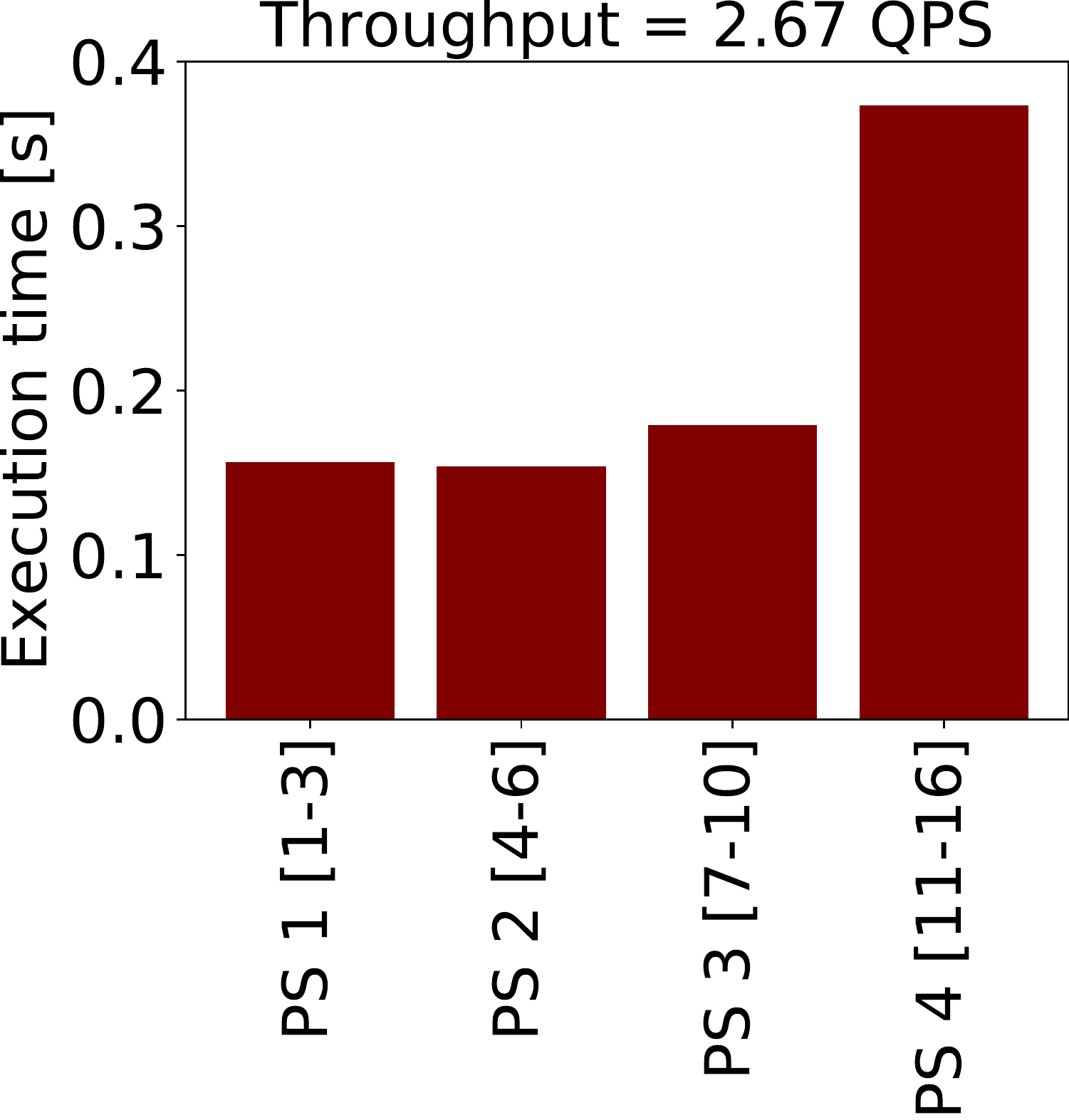}
        \caption{Interference on stage 4}
        %Execution on 4 execution places with interference on stage 4
        \label{fig:motivation_2_case_2}
  \end{subfigure}
\end{minipage}
\hspace{0.3pt}
\begin{minipage}[t]{.22\linewidth}
  \centering
  \begin{subfigure}[t]{\textwidth}
       \includegraphics[width=0.99\linewidth]{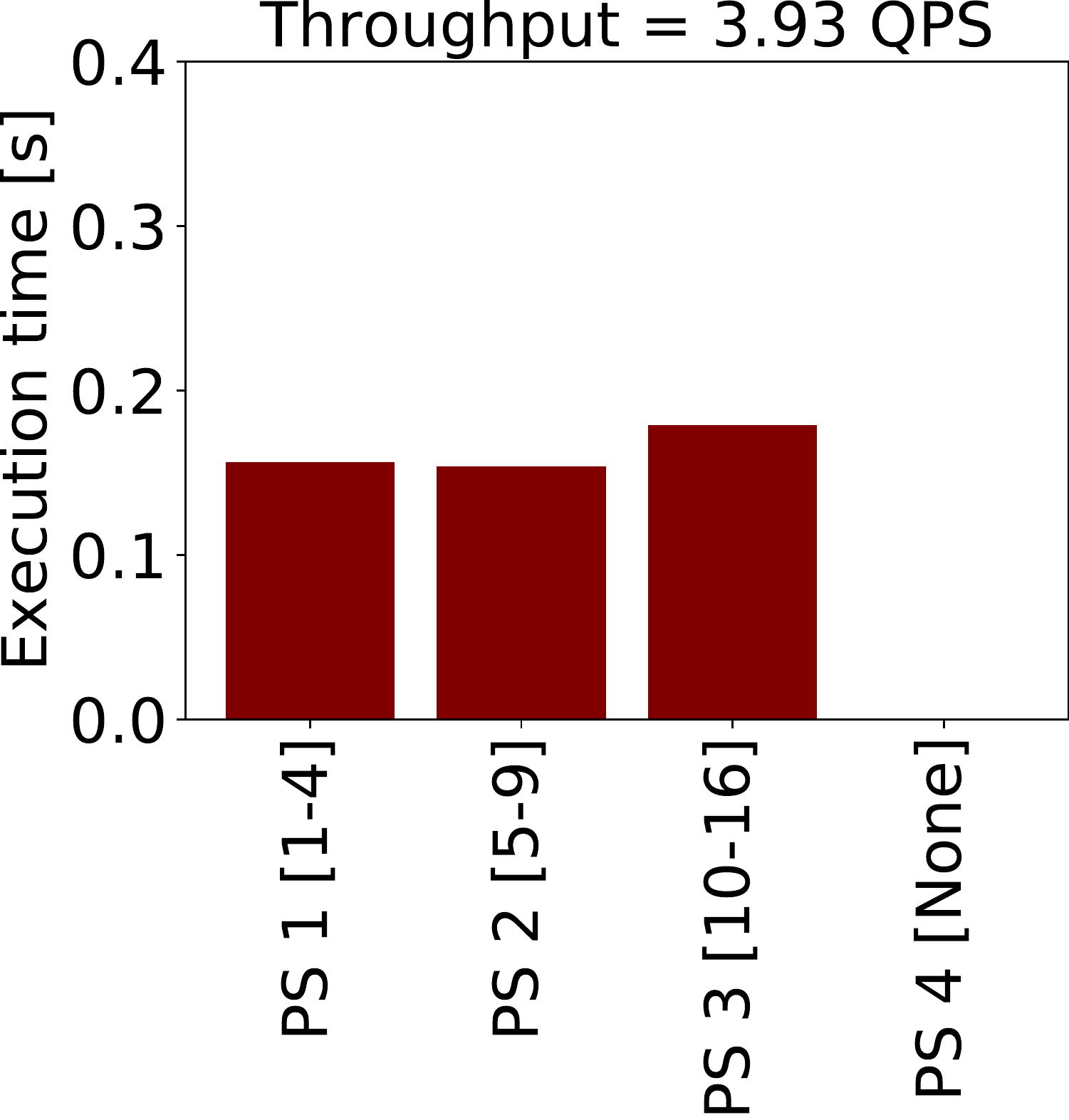}
        \caption{Execution with 3 stages}
        %Interference mitigation by execution on interference-free resources with 3 stages.
        \label{fig:motivation_2_case_3}
  \end{subfigure}
\end{minipage}
\hspace{0.3pt}
\begin{minipage}[t]{.22\linewidth}
  \centering
  \begin{subfigure}[t]{\textwidth}
       \includegraphics[width=0.99\linewidth]{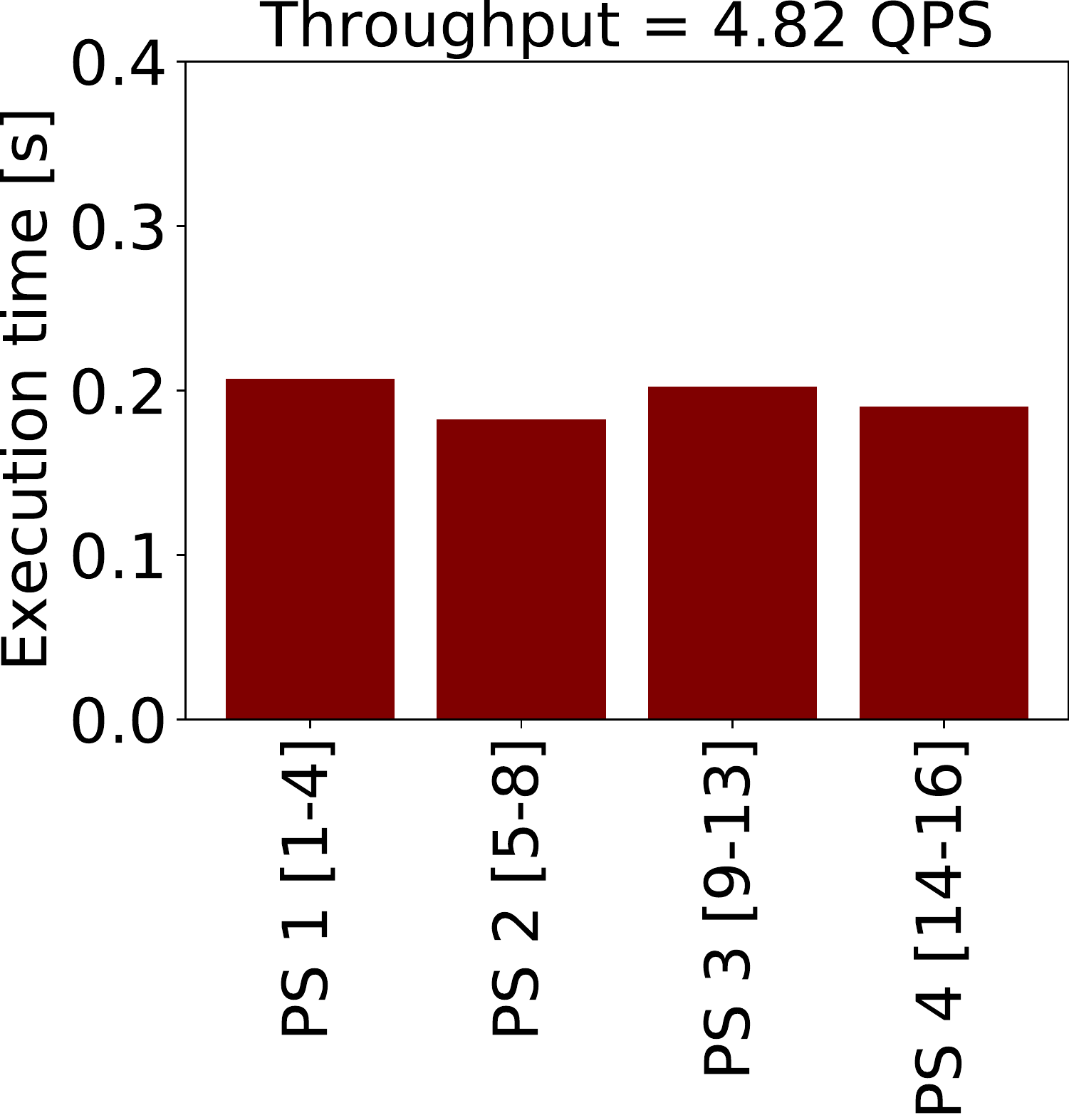}
        \caption{Exhaustive search rebalancing}
        %Interference mitigation after pipeline rebalancing through exhaustive search.
        \label{fig:motivation_2_case_4} 
  \end{subfigure}
\end{minipage}
\caption{Throughput and execution time of a 4-stage pipeline for VGG-16. }
\label{fig:motivation_2}
\end{figure}

Figure~\ref{fig:motivation_2} shows a motivating example of an inference pipeline for VGG16, a CNN model. The pipeline consists of 4 stages, each consisting of 3 to 5 layers of the network model (Figure~\ref{fig:motivation_2_case_1}), in a configuration where the pipeline stages are balanced in terms of execution time. Assuming a workload is colocated on the execution place which executes the fourth stage of the pipeline, the execution time of this stage increases due to interference, causing the throughput to decrease by 46\% (Figure~\ref{fig:motivation_2_case_2}). A static solution would dedicate the resources to the colocated workload, and would use only 3 execution places. To maintain high throughput, the pipeline stages would also be reduced to 3, leading to a suboptimal solution (Figure~\ref{fig:motivation_2_case_3}). A dynamic solution would attempt to rebalance the initial four pipeline stages, to mitigate the effect of interference on the execution time of the fourth stage. An exhaustive search for an optimal new configuration is able to restore the initial throughput loss (Figure~\ref{fig:motivation_2_case_4}), however this exhaustive search required $42.5$ minutes to complete. % \textcolor{red}{took} $42.5$ minutes to complete.   
%Figure\ref{fig:motivation_2} shows the execution times of 4-stage pipeline for VGG16. An interference benchmark is co-scheduled on the resources which are already assigned to stage 4 of the pipeline, this causes $46\%$ degradation in the pipeline throughput. Figure \ref{fig:motivation_2_case_3} shows a new configuration of pipeline in the presence of interference. The new configuration is found after exhaustive search of all possible configurations which takes $42.5$ minutes to complete. Another solution could be to re-schedule the pipeline by delegating all the resources of stage 4 to co-scheduled application as shown in Figure \ref{fig:motivation_2_case_4} leading to a sub-optimal solution.

This experiment allows us to make the following observations: First, the effect of interference on a parallel inference pipeline can be mitigated by rebalancing the pipeline stages. Second, partitioning the resources between the colocated workload and the inference pipeline leads to a shorted pipeline and a suboptimal throughput. Third, dynamic reaction to interference is able to largely restore throughput loss on the inference pipeline. Fourth, an exhaustive search for an optimal configuration is infeasible in a reactive, dynamic solution. The above observations motivate our work, which proposes an online scheduling technique for the pipeline stages of inference pipelines. 

%\textbf{Our solution dynamically detects changes in the pipeline throughput due to interference and uses a heuristic to find a near-optimal configuration of pipeline stages, which avoids interference and restores throughput in feasible time.} 
 
% Figure \ref{fig:pipeline_motivation} futher shows that in an attempt to re-balance the pipeline, we move layers among the pipeline stages to improve the performance even in the presence of interference. These observations motivates the automation of rebalancing the inference pipeline in presence of interference in a faster way, eliminating the need to migrate the whole application or re-running the scheduler to find a new assignment for both co-running applications.  
% \begin{figure}[!tb]
% \includegraphics[width=\linewidth]{shisha/figs/motivation_1.pdf}
% \caption{Impact of interference on a convolutional layer and re-adjustment of resources (8 cores) among two kernels. A corresponds to a compute intensive CNN layer and B corresponds to interference benchmark stressing CPUs}\label{fig:resnet50_motivation}
% \end{figure}

%% file: shisha/background.tex
%\section{Background}
  

%% file: shisha/methodology.tex
\section{ODIN: A dynamic solution to overcome interference on inference pipelines}
\subsection{Methodology}
\begin{figure}[!t]
\centering
\includegraphics[width=0.8\linewidth]{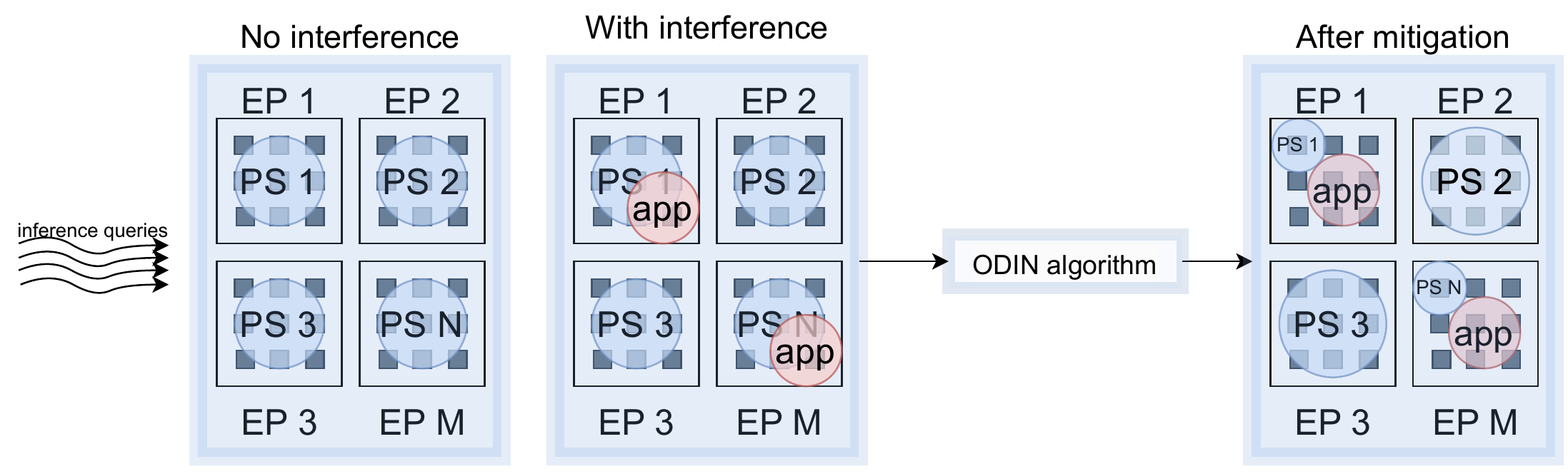}
\caption{System overview}\label{fig:system_overview}
\end{figure}
In this work, we consider a system with a set of resources named execution places (EPs). Each execution place may consist of multiple cores, but execution places do not share performance-critical resources between them, e.g. caches, memory controllers/links. 
%\textbf{Although this work focuses on multicore systems with multiple sockets, this system abstraction is applicable to other types of systems, as are, for example, multi-GPU nodes, \nikela{where, however, ODIN would additionally have to take into account the transfer times of layers between the different devices}. }%\textcolor{red}{in which execution times of kernels optimized specifically for GPUs could be used as input to ODIN}. 
Inference pipelines are linear and are implemented with a bind-to-stage approach, where a single pipeline stage
 (PS) is assigned to a single EP, i.e. a unique set of resources of the system, and pipeline stages do not share resources. Pipeline stages can exploit the multiple resources within an EP by other means of parallelism, e.g. operator parallelism. A pipeline configuration defines the mapping of pipeline stages to execution places and the assignment of layers of a neural network model to PSs. We additionally assume that, in an interference-free system where the inference pipeline utilizes all the available execution places, the stages are already effectively balanced across the execution places. 
%\textcolor{red}{across the execution places}, \nikela{by some orthogonal balancing technique for inference pipelines}. 
If a workload is colocated with a pipeline stage on one of the EPs, causing interference and increase of the execution time of this stage, the heuristics which form the backbone of our solution attempts to reduce the total work on the affected pipeline stage, moving network layers to non-affected pipeline stages.
%new
A high-level overview of our approach, ODIN, is presented in Figure~\ref{fig:system_overview}. Our approach operates online and is agnostic to any other colocated application. At runtime, we monitor the execution time of pipeline stages, and scan for changes in the performance of the slowest pipeline stage. If its execution time has increased, we consider it as affected by an interfering application and trigger the online rebalancing of pipeline stages, to find a new configuration, using our heuristic algorithm. If its execution time has decreased, we consider that any effect of interference is no longer present, and once again trigger online rebalancing to find a new configuration that reclaims resources from the colocated, interfering workload.\looseness=-1 
%At runtime, we monitor the execution time of bottleneck stage since the performance of a PS is determined by the bottleneck, any change (positive or negative) in its performance calls for re-balancing of the pipeline stage. Therefore when a co-scheduled application ends and inference pipeline reclaims the resources, the pipeline configuration might not be suitable for the new state of the system. We run our mitigation algorithm again to find a new pipeline configurati
\subsection{ODIN: A heuristic-based approach for pipeline stage re-balancing under interference}

\begin{algorithm}[!tbh]
\footnotesize 
\begin{algorithmic}[1]
    \Require{$C$, $\alpha$} \Comment{$C$ = pipeline configuration}
    
    \State $T \gets \textproc{throughput}(C)$ \Comment{$T$ = throughput of the pipeline}
    \State $C_{\textit{opt}} \gets C$ \Comment{Optimal pipeline configuration}
    \State $\gamma \gets 0$ \Comment{counter variable}
    
    \While{$\gamma < \alpha$}
        \State $\textnormal{PS}_\textit{affected} \gets \textproc{get\_index}(\max(t(C)))$
        
        \If{$\gamma = 0$}
            \State $C[\textnormal{PS}_{\textit{affected}}+1] \mathrel{+}= 1$
            \State $C[\textnormal{PS}_{\textit{affected}}-1] \mathrel{+}= 1$
            \State $C[\textnormal{PS}_{\textit{affected}}] \mathrel{-}= 2$
        \EndIf
        
        \State $\textnormal{S}_{\textit{left}} \gets \textproc{sum}(t(C[0], C[\textnormal{PS}_{\textit{affected}}]))$
        \State $\textnormal{S}_{\textit{right}} \gets \textproc{sum}(t(C[\textnormal{PS}_{\textit{affected}}+1], C[N]))$
        
        \If{$\textnormal{S}_{\textit{left}} < S_{\textit{right}}$}
            \State $direction \gets \textit{left}$
        \Else
            \State $direction \gets \textit{right}$
        \EndIf
        
        \State $\textnormal{PS}_{\textit{lightest}} \gets \textproc{get\_index}(t(C, \textnormal{PS}_{\textit{affected}}, direction))$
        \State $C[\textnormal{PS}_{\textit{affected}}] \mathrel{-}= 1$
        \State $C[\textnormal{PS}_{\textit{lightest}}] \mathrel{+}= 1$
        \State $\textnormal{T}_{\textit{new}} \gets \textproc{throughput}(C)$
        
        \If{$\textnormal{T}_{\textit{new}} < \textnormal{T}$}
            \State $\gamma \mathrel{+}= 1$
        \ElsIf{$\textnormal{T}_{\textit{new}} = \textnormal{T}$}
            \State $C[\textnormal{PS}_{\textit{affected}}] \mathrel{-}= 1$
            \State $C[\textnormal{PS}_{\textit{lightest}}] \mathrel{+}= 1$
            \State $\gamma \mathrel{+}= 1$
        \Else
            \State $\gamma \gets 0$
            \State $\textnormal{T} \gets \textnormal{T}_{\textit{new}}$
            \State $C_{\textit{opt}} \gets C$
        \EndIf
    \EndWhile     \\
\Return $C_{\textit{opt}}$
\end{algorithmic}

    \caption{ODIN Algorithm}
    \label{alg:tuning}

\end{algorithm}

We describe our approach, ODIN, to mitigate the effect of interference on parallel inference pipelines, and the heuristics it uses to find new configurations for the pipeline stages at runtime. The complete steps of our approach are presented in Algorithm~\ref{alg:tuning}. The algorithm takes as input the current configuration $C$, which tracks the number of network layers belonging to each pipeline stage, and a tuning parameter $\alpha$. As the algorithm starts operating without interference, the current configuration is considered to be optimal, and the pipeline throughput is the one given by the current configuration. During execution, the execution time of PSs is monitored.  Interference is detected when the execution time $t$ of one of the pipeline stages increases. We identify the affected PS ($\textnormal{PS}_{\textit{affected}}$) as the slowest stage in the current configuration, and this determines the throughput of the pipeline. The goal of the algorithm is then to rebalance the pipeline stages by removing layers from the affected PS, to reduce its work. We note that, removing layers from the affected PS may reduce the length of the pipeline by 1. We apply two heuristics to find a new configuration: 
\\
\textbf{1) Set the direction for moving work: } 
To remove layers from the affected PS, we first determine the direction of moving the layers. As the layers of an inference pipeline execute one after the other (forward pass), we can only remove layers from the head or tail of the $\textnormal{PS}_{\textit{affected}}$. At the first attempt, the algorithm does not know which layers of the $\textnormal{PS}_{\textit{affected}}$ have experienced performance degradation due to interference, so we initially remove layers from both ends, as shown in Lines 6-10, and move them to the preceding and subsequent pipeline stages respectively. Next, we calculate the sum of the execution time of PSs on both sides of the $\textnormal{PS}_{\textit{affected}}$ and set the direction to move layers. We then find the PS with the lowest execution time $\textnormal{PS}_{\textit{lightest}}$ in that direction, starting from $\textnormal{PS}_{\textit{affected}}$, and move one layer to $\textnormal{PS}_{\textit{lightest}}$, as shown in Lines 18-20. \looseness=-1
\\
\textbf{2) Avoiding Local optimum }
Our first heuristic may result in a local, rather than a global optimum. A possible solution for this is to randomly choose a completely new starting configuration, and rebalance again. However, this can lead to loss of information. Since our initial configuration is optimal for the execution of the pipeline in an interference-free case, in the case of a local optimum, we deliberately move more layers from the $\textnormal{PS}_{\textit{affected}}$ to the $\textnormal{PS}_{\textit{lightest}}$, to create a different configuration and continue the exploration.

The extent of exploration is controlled by variable $\alpha$ which is provided as an input to the algorithm. As the algorithm is applied online, while the inference pipeline is running, the value of $\alpha$ can be tuned to reduce the number of trials for faster exploration. 
\begin{figure}[t]
\centering
\includegraphics[width=0.9\linewidth]{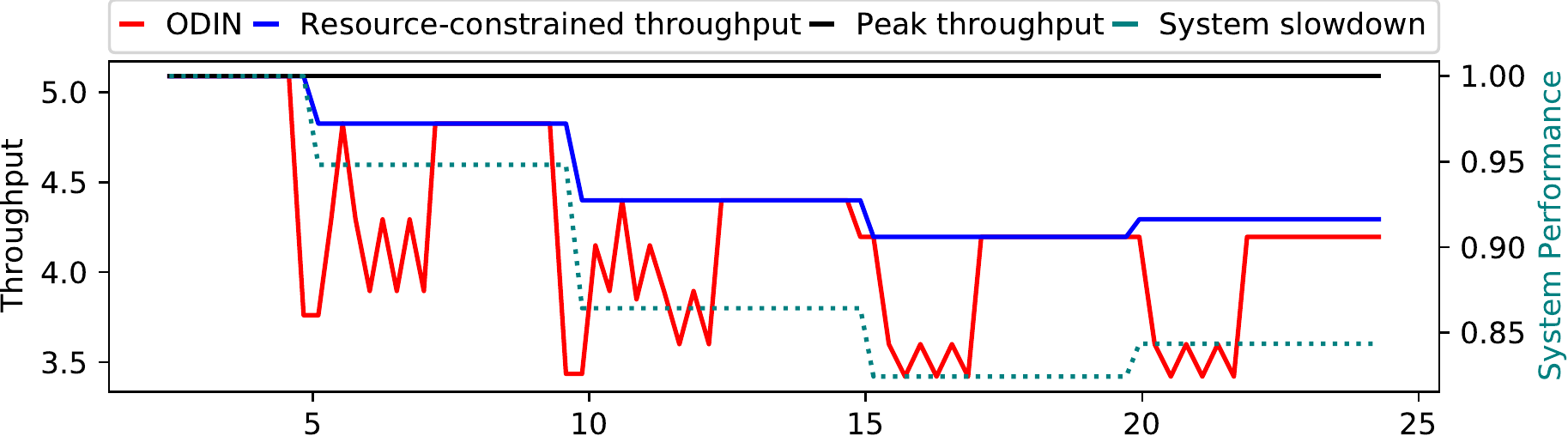}
\caption{A timeline of a VGG16 inference pipeline, running with ODIN, which reacts to mitigate interference at time steps 5, 10, 15, and 20.}\label{fig:timeline}
\end{figure}
Figure~\ref{fig:timeline} shows a timeline of an inference pipeline for VGG16, executing on four EPs with pipeline stages, where ODIN runs to mitigate the effects of interference. Initially, there is no interference, the inference pipeline is balanced with an optimal configuration and achieves its peak throughput. At time steps 5, 10, and 15, a new workload is co-located on a different execution place, slowing down the system for the inference pipeline, reducing what we define as the resource-constrained throughput, i.e.~the throughput the inference pipeline can attain in the presence of interference. At each of these time steps, ODIN automatically detects the throughput degradation and rebalances the pipeline until it finds a successful solution. At time step 20, one of the interfering workloads is removed, and ODIN executes again, to restore the pipeline throughput by claiming back the resources previously used by the colocated workload. \looseness=-1
\subsection{Implementation details}
\subsubsection{Database Creation}: 
In our evaluation, we use simulation to be able to apply ODIN on any type and size of the underlying system. We, therefore, replace online monitoring with an offline database. We first collect the execution time of the $m$ individual network layers of the inference pipelines under consideration, when executing alone (without any interference), on a real platform. On the same platform, we collect the execution time of the individual network layers when executing alongside co-located applications, producing $n$ different interference scenarios. We then store these collected $m \times (n+1) $ measurements in a database, and use them in simulation. We consider the real platform to be a single execution place for ODIN, and simulate multiple execution places of the same type. To emulate interference, during simulation, we randomly select an interference scenario for an execution place and look up the corresponding execution time in the database.
%We, therefore, replace online monitoring with an offline database, where we store the execution time of individual network layers of the inference pipelines we execute, on the execution places of the platform we simulate. For each layer, our database stores its execution time when executing alone (without any interference) and when executing alongside a co-located application, which creates interference.
%For a \textcolor{red}{network} pipeline of $m$ layers, our database consists of $m \times (n + 1)$ execution time measurements, where $n$ is the number of different colocated, interfering applications we examine. \textcolor{red}{The entries of the database are recorded by running the networks layers on our testing platform.} \looseness=-1 
\subsubsection{Throughput calculation}: 
We use the measurements in our database $D$ of size $m \times n$ to calculate the throughput of a pipeline, as follows: 

\begin{center}
\vspace{-2mm}
 $T = \frac{1}{max_{i = 0}^N \sum_{l=0}^{P} D[l,k]}  $ 
 \vspace{-2mm}
\end{center}
\noindent where $N$ is the number of pipeline stages, $P$ is the number of layers in a pipeline stage, and $D[l,k]$ is the execution time of layer $l$ under the type of interference $k$, as recorded in the database $D$.
%To work with an inference serving system we simulated the environment by creating database of execution times of individual network layers. For each layer we profiled the execution time in interference and non-interference scenario. For any type of co-scheduled application, we run network layers along with interference applications and profile the timing. We record the timings in a database and use this database to test different combinations of layers per stage to calculate the throughput of a pipeline configuration.
%The pipeline throughput from the database is calculated from Equation~\ref{eq1:database_eq}. Where, $N = $ Number of pipeline stages, $P = $Number of layers within a pipeline stage, $D$ is a $2D$ matrix in which $l$ represents layer number in the network and $k$ represents interference scenario. We create database D of size $m\times n$ for every network, m is the total number of layers in a network, n is the number of interference scenarios we consider for the simulations. 

\subsubsection{Implementation of the least-loaded scheduler (LLS) as a baseline}: 
LLS is an online interference mitigation technique~\cite{delimitrou2013paragon, devi2016load, shaw2014survey}. We implement LLS in the context of pipeline stages, as a baseline to compare against ODIN. We calculate the utilization of each pipeline stage and move the layers from the most utilized to the least utilized stage recursively until the throughput starts decreasing. The utilization of a stage $\upsilon_i$ is calculated as: 
\begin{center}
\vspace{-2mm}
$ \upsilon_i = \left ( 1-\frac{w_i}{w_i + t_i} \right )  $
\vspace{-2mm}
\end{center} 

\noindent where $t_i$ is the execution time of a pipeline stage, and $w_i$ is the waiting time of the stage, calculated as $w_i = w_{i-1} + t{i-1} - t_i$, with $w_0 = 0$. \looseness=-1 
% \begin{table}[]
% \centering
% \begin{tabular}{|c|c|c|}
% \hline
% Throughput                                         & Utilization                                              & Waiting time                                                                                                    \\ \hline
% $T = \frac{1}{max_{i=0}^N \sum_{l=0}^{P} D[l,k]}$ & $\upsilon_i = \left ( 1-\frac{w_i}{w_i + t_i} \right ) $ & $W_i = \left\{\begin{matrix}w_{i-1} + t_{i-1} - t_i&\text{, } i \neq 0 \\ 0&\text{, } i = 0\end{matrix}\right.$ \\ \hline
% \end{tabular}
% \end{table}

%% file: shisha/evaluation.tex
  %\begin{minipage}[h]{\textwidth}
  \begin{minipage}[h]{0.45\textwidth}
    %\centering
    \includegraphics[width=0.88\linewidth]{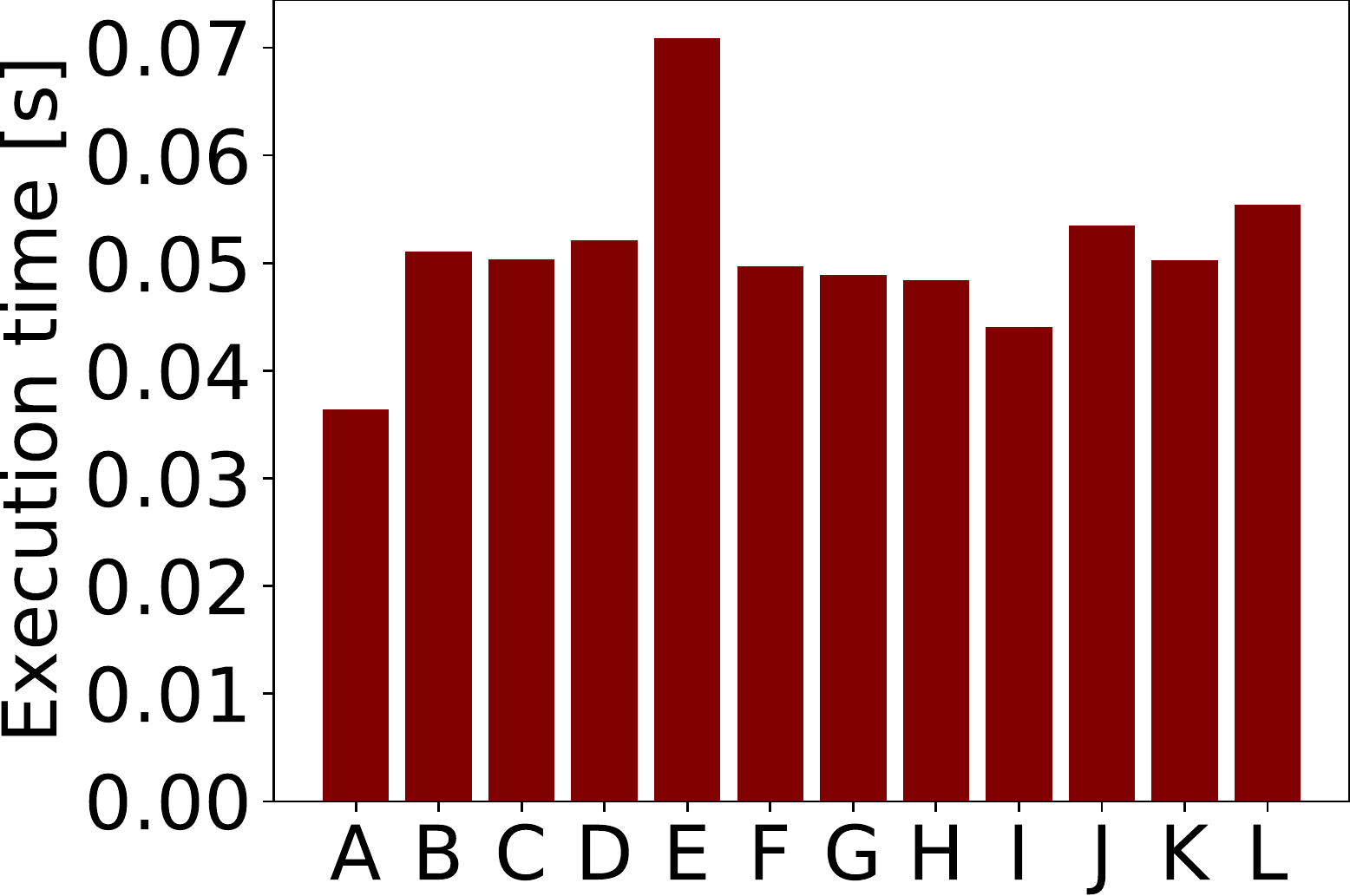}
    \captionof{figure}{Performance impact} %of a single vgg16 layer executed on 8xP-Cores of Alder Lake
    \label{fig:experimental_setup}
  \end{minipage}
  %\hfill
  \begin{minipage}[h]{0.45\textwidth}
    \centering
    \resizebox{1.1\linewidth}{!}{%
\begin{tabular}{@{}ll@{}}
\toprule
Mode of execution & Core assignment on Alder Lake                                \\ \midrule
A                 & CNN:{[}0-7{]}                                                \\
B                 & CNN: {[}0-7{]}, IBench-MemBW: {[}0{]}                        \\
C                 & CNN: {[}0-7{]}, IBench-MemBW: {[}0-1{]}                      \\
D                 & CNN: {[}0-7{]}, IBench-MemBW: {[}0-3{]}                      \\
E                 & CNN: {[}0-7{]}, IBench-MemBW: {[}0-7{]}                      \\
F                 & CNN: {[}0-7{]}, IBench-CPU: {[}0{]}                          \\
G                 & CNN: {[}0-7{]}, IBench-CPU: {[}0-1{]}                        \\
H                 & CNN: {[}0-7{]}, IBench-CPU: {[}0-3{]}                        \\
I                 & CNN: {[}0-7{]}, IBench-CPU: {[}0-7{]}                        \\
J                 & CNN: {[}0-3{]}, IBench-MemBW: {[}4-7{]}                      \\
K                 & CNN: {[}0-3{]}, IBench-CPU: {[}4-7{]}                        \\
L                 & CNN: {[}0-3{]}, IBench-CPU: {[}4-7{]}, IBench-MemBW{[}4-7{]} \\
\bottomrule
\end{tabular}}
      \captionof{table}{Interference scenarios}
        \label{tab:interference_scenarios}
    \end{minipage}
  %\end{minipage}

\section{Evaluation}
\subsection{Experimental setup}
We execute ODIN in a simulated system for inference serving, which consists of multiple execution places, and each execution place consists of a fixed number of 8 cores. To generate our database, we use an Intel i9-12900K (AlderLake) server, which consists of $8$ $2$xP-cores (Performance) and $8$ $2$xE-cores (Efficient). We consider the set of 8 P-cores as a single execution place in our system. 

For the neural network models we examine as inference pipelines, our database consists of measurements for each layer without interference, as well as measurements for each layer with 12 different co-located workloads, in different settings. To create the co-located workloads, we use two interference benchmarks from the \texttt{iBench} suite ~\cite{delimitrou2013ibench}, the \texttt{CPU} benchmark that stresses the CPU and the \texttt{memBW} benchmark that stresses the memory bandwidth. We then create our 12 scenarios of colocation by assigning the network layers and interference benchmarks different numbers of threads, and pinning them to different cores. Table~\ref{tab:interference_scenarios} showcases the colocation scenarios considered in our database, and Figure~\ref{fig:experimental_setup} demonstrates the performance impact of interference for all these colocation scenarios on a single layer of the VGG16 network model. 

For our evaluation, we consider the inference pipelines of three popular CNN models: VGG16 \cite{simonyan2014very}, ResNet-50 and ResNet-152 \cite{he2016deep}, with 16, 50, and 152 layers respectively, implemented with the Keras \cite{chollet2015keras} framework.
\subsection{Interference mitigation with ODIN}
%experimental setup
To evaluate the effectiveness of ODIN, we compare its latency and throughput for different values of $\alpha$, which sets the extent of exploration, against LLS, in several interference scenarios. In particular, we consider a system of 4 executions places of 8 cores each, which serves inference queries with two network models, VGG16, and ResNet-50. We assume a fixed number of 4000 queries, and induce random interference on different execution places, based on the colocation scenarios described in Table~\ref{tab:interference_scenarios}. We consider different values for the frequency (frequency periods of 2, 10, and 100 queries) and duration (2, 10, and 100 queries) of interference, and evaluate the end-to-end latency and throughput distribution of each inference pipeline.   \looseness=-1 %Latency

\begin{figure}[!tb]

    \begin{subfigure}[]{.44\textwidth}
        \includegraphics[width=\linewidth]{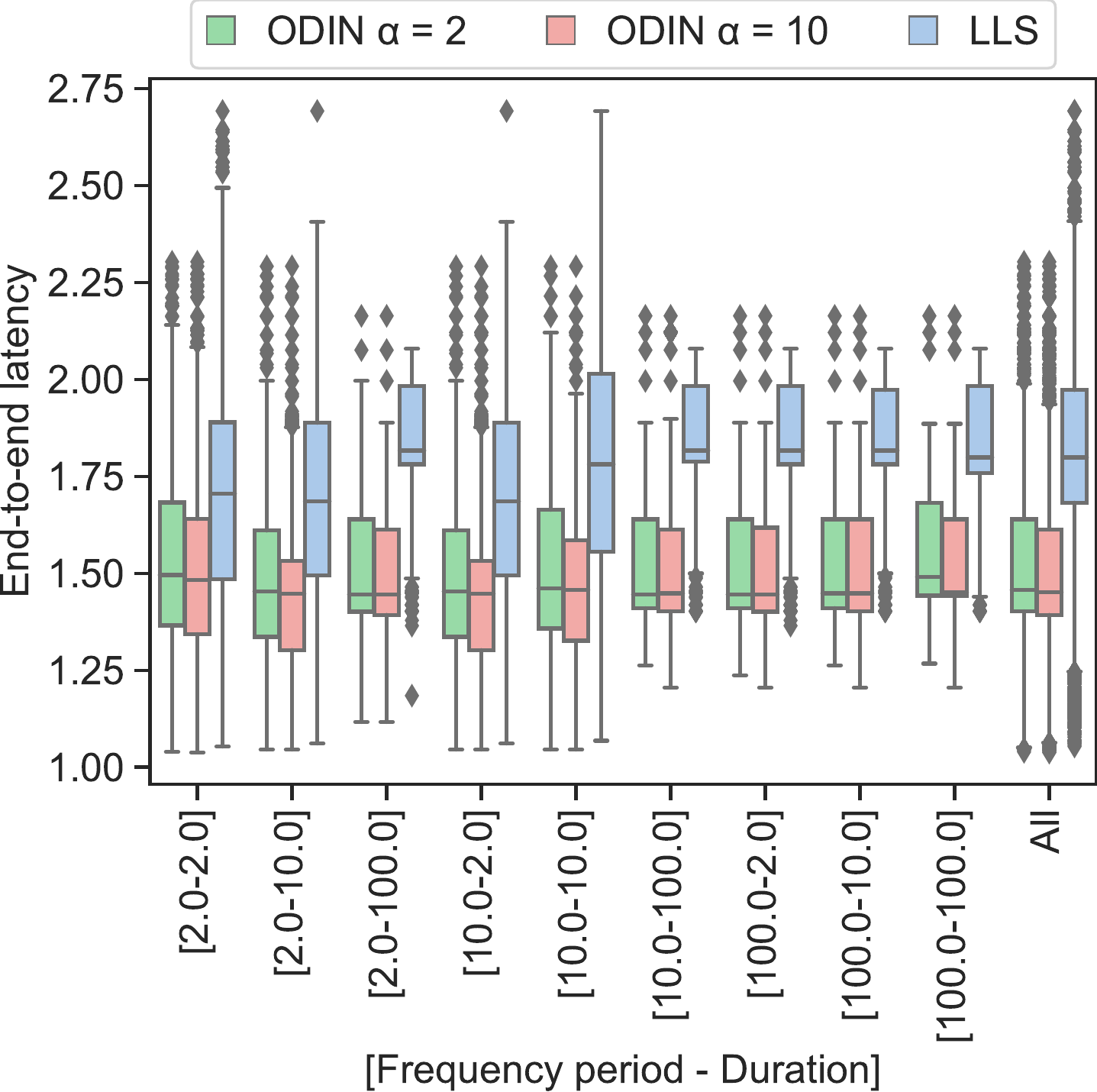}
        \caption{ResNet50}
        \label{fig:lat_resnet}
    \end{subfigure}
\hfill
  \begin{subfigure}[]{.44\textwidth}
        \includegraphics[width=\linewidth]{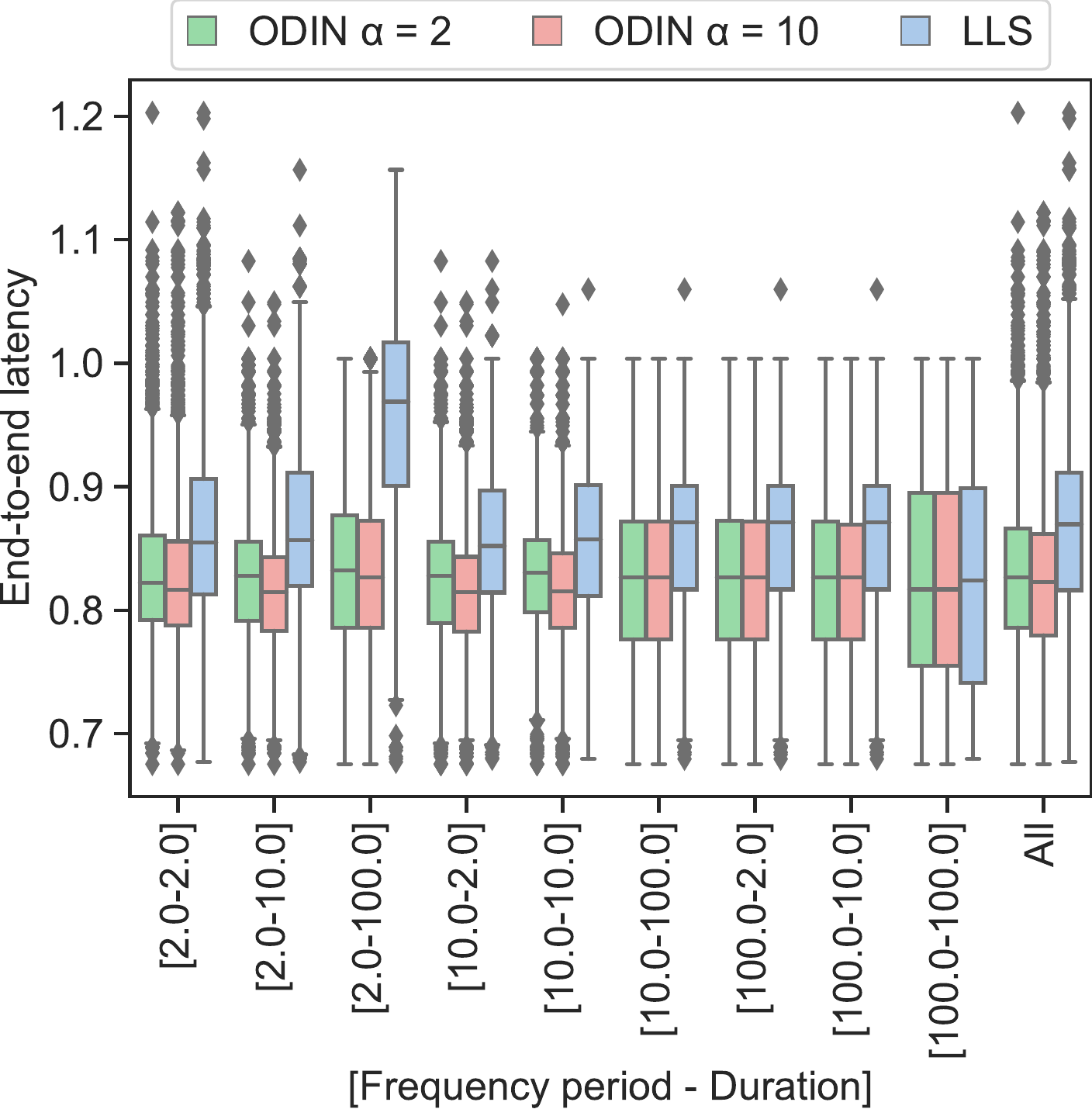}
        \caption{VGG16}
        \label{fig:lat_vgg16}
  \end{subfigure}

\caption{Inference pipeline latency (lower is better) with ODIN, in comparison to LLS, over a window of 4000 queries, for interference of different frequency period and duration. \looseness=-1 }
\label{fig:lat_resnet_vgg16}
\end{figure}

\noindent \textbf{Latency:} Figure~\ref{fig:lat_resnet_vgg16} shows the latency distribution of the two inference pipelines under interference. 
%Observation 1
We observe that ODIN outperforms LLS in all scenarios, delivering lower latency. We highlight the effect of the $\alpha$ parameter of ODIN on latency. A higher value of $\alpha$ yields lower latency, because the longer exploration phase allows ODIN to find an optimal configuration. On the other hand, if the frequency of interference is high, a low value of $\alpha$ is able, in most cases, to produce an equally good solution with lower exploration time.  
%Observation 2
ODIN $\alpha = 10$ yields better latency than ODIN $\alpha = 2$ this is because the former takes more trials to find a schedule, however if the frequency of interference is high then it may take longer to find a solution or end up with sub-optimal solution. We additionally note that both ODIN and LLS are more effective in cases where interference appears with lower frequency and for longer periods. This is particularly evident in Figure~\ref{fig:lat_resnet_vgg16}. For the pair of [frequency period = 2, duration = 2], the distribution of latency shows many outliers, as an optimal configuration found by the algorithm for one period of interference may be applied to the next period, where the pattern of interference has changed. 
Overall, however, ODIN outperforms LLS in all scenarios, offering 15.8\% better latency on average with $\alpha=10$ and 14.1\% with $\alpha=2$. \looseness=-1
%Throughput

\begin{figure}[!tb]
\centering

    \begin{subfigure}[]{.44\textwidth}
        \includegraphics[width=\linewidth]{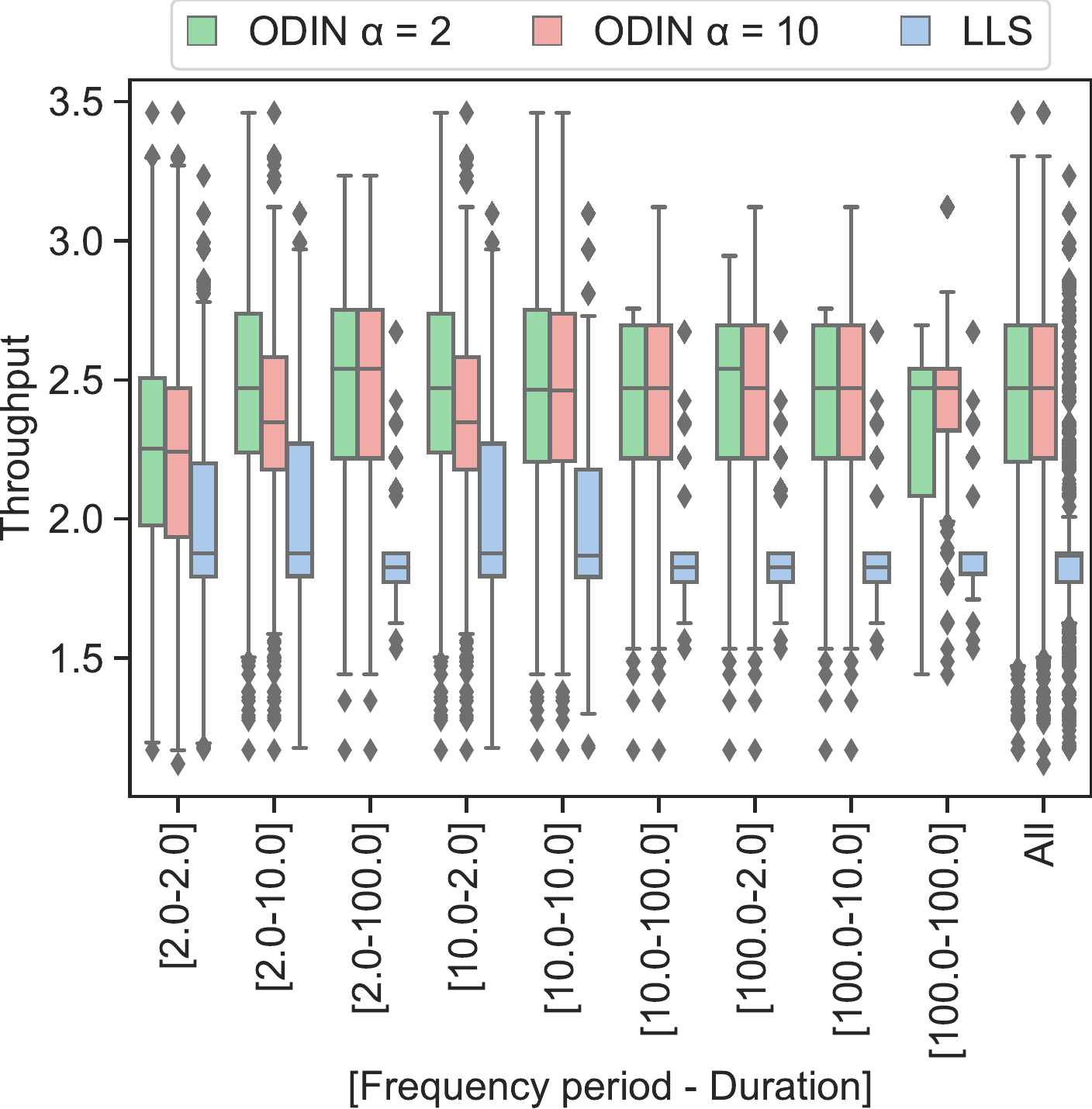}
        \caption{ResNet50}
        \label{fig:tp_resnet}
    \end{subfigure}
\hfill
  \begin{subfigure}[]{.44\textwidth}
        \includegraphics[width=\linewidth]{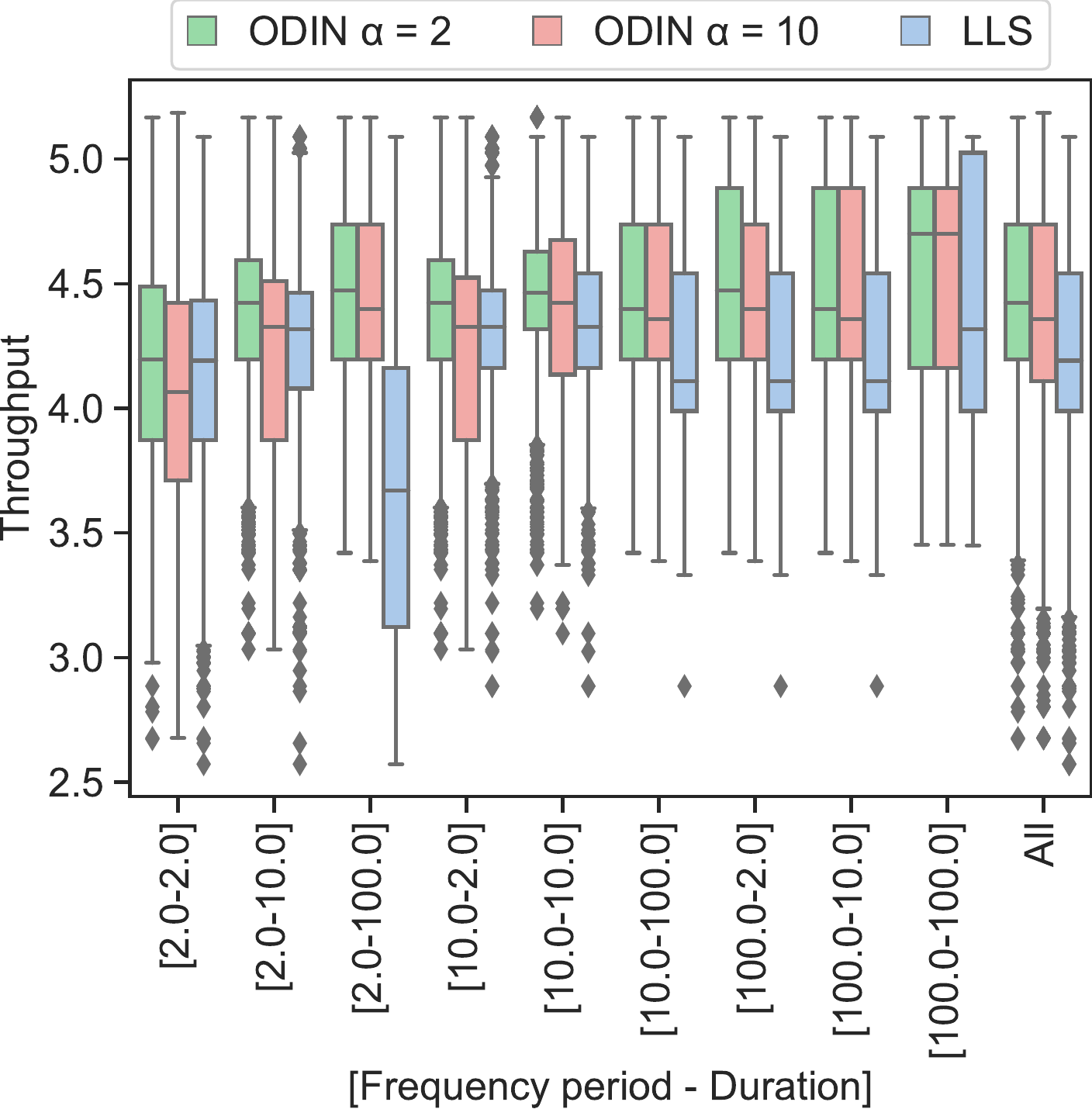}
        \caption{VGG16}
        \label{fig:tp_vgg16}
  \end{subfigure}

\caption{Inference pipeline throughput (higher is better) with ODIN, in comparison to LLS, over a window of 4000 queries, for interference of different frequency period and duration. \looseness=-1 }
\label{fig:tp_vgg16_resnet}
\end{figure}

\noindent \textbf{Throughput:} We then compare the throughput of the inference pipelines under interference, for ResNet50 and VGG16, with ODIN and LLS, for the same interference scenarios, in Figure~\ref{fig:tp_vgg16_resnet}. Again, ODIN offers higher throughput than LLS in most cases. The case of VGG16 highlights our observation about the lower performance in the case of high frequency, where all three techniques show outliers of low throughput, however, ODIN is more able to adapt to interference of longer duration compared to LLS. We observe additionally that for the case of the highest frequency period-duration pair [100, 100], LLS and ODIN have comparable performance, as the near-optimal solutions were obtained with minimal changes of the pipeline configurations. Overall, on average, ODIN achieves 19\% higher throughput than LLS with any choice of $\alpha$. \looseness=-1

\begin{figure}[!tb]

%\begin{minipage}[]{.49\textwidth}
%    \centering
    \begin{subfigure}[]{.42\textwidth}
        \includegraphics[width=\linewidth]{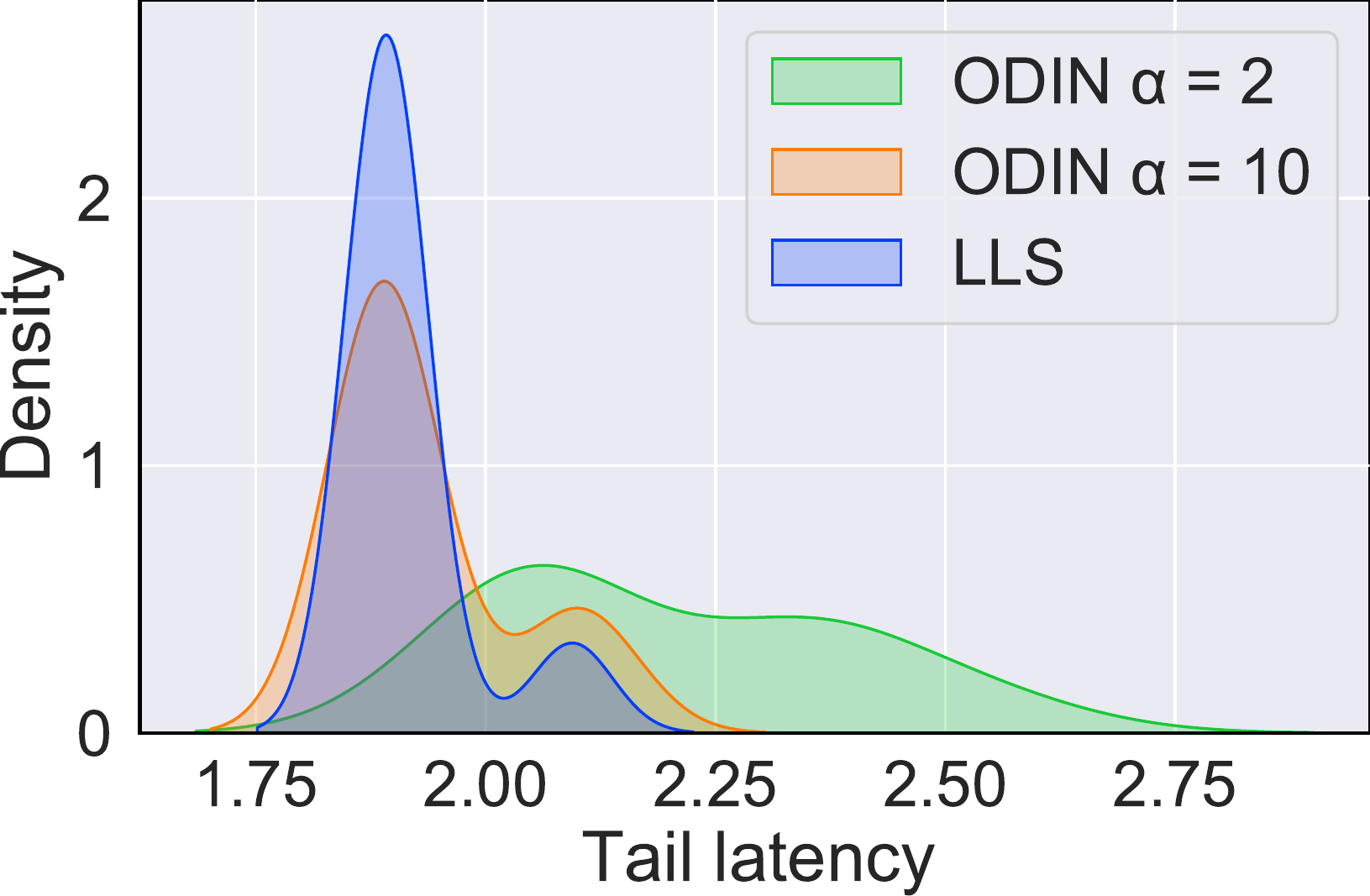}
        \caption{ResNet50}
        \label{fig:lat_resnet_dist}
    \end{subfigure}
%\end{minipage}
%\hspace{0.3pt}
%\begin{minipage}[]{.49\linewidth}
%  \centering
\hfill
  \begin{subfigure}[]{.42\textwidth}
        \includegraphics[width=\linewidth]{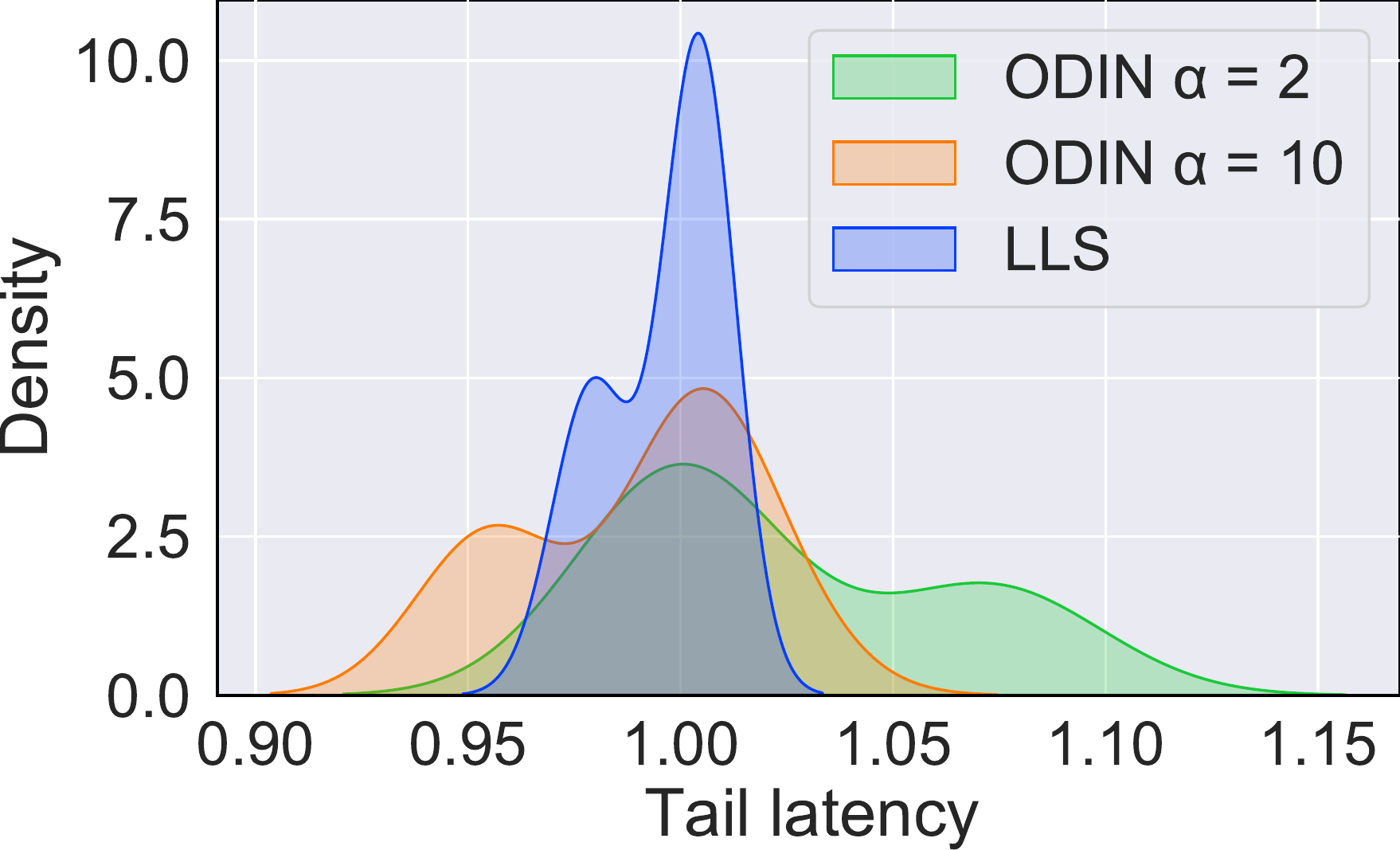}
        \caption{VGG16}
        \label{fig:lat_vgg16_dists}
  \end{subfigure}
%\end{minipage}
\caption{Tail latency distribution of ODIN, in comparison to LLS.}
\label{fig:lat_vgg_renet_dist}
\end{figure}

%Observation 3
%In the case high frequency in interference such as [Frequency period - Duration] = [2.0-2.0] in Figure ~\ref{fig:lat_resnet_vgg16} The state of the systems changes frequently, the decision taken by the algorithm for one state gets applied on a different scenario, this results in higher latency values. On the other side cases when interference runs for longer time such as [Frequency period - Duration] = [100.0-100.0], the distribution of latency less dense. Note that the latency distribution also depends on random interference scenarios induced in the system, the effect on performance is higher in one scenario than another. The bottomline for this analysis is that on average ODIN performs 15\% better than LLS in all cases. 
\noindent \textbf{Tail latency:} Besides the latency distribution, we separately examine the tail latency (99th percentile), as it can be a critical metric in inference-serving systems, and it is also indicative of the quality of the solutions found by ODIN. Figure~\ref{fig:lat_vgg_renet_dist} shows the distribution of the tail latency across all the queries considered in the interference scenarios examined in this Section. For both ResNet50 and VGG16, ODIN results in significantly lower tail latencies than LLS. For the case of VGG16, we additionally observe that a higher value of $\alpha$ for ODIN can produce better solutions, resulting in lower tail latencies. On average, ODIN results in 14\% lower tail latencies than LLS. \looseness=-1

\begin{figure}[!tb]

%\begin{minipage}[]{.49\textwidth}
%    \centering
    \begin{subfigure}[]{0.42\textwidth}
        \includegraphics[width=0.95\linewidth]{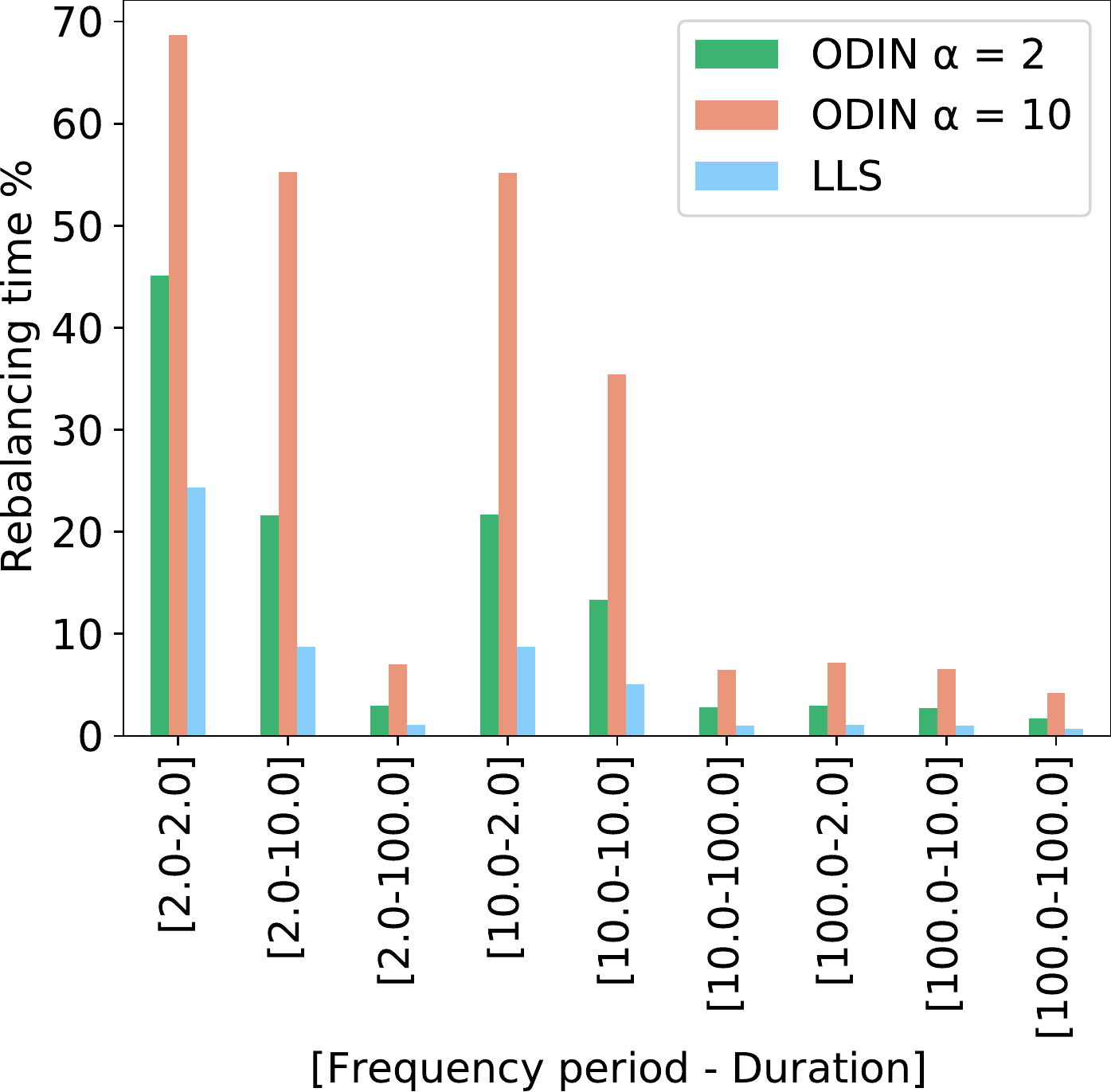}
        \caption{ResNet50}
        \label{fig:ov_resnet}
    \end{subfigure}
%\end{minipage}
\hfill
%\begin{minipage}[]{.49\linewidth}
%  \centering
  \begin{subfigure}[]{0.42\textwidth}
        \includegraphics[width=0.95\linewidth]{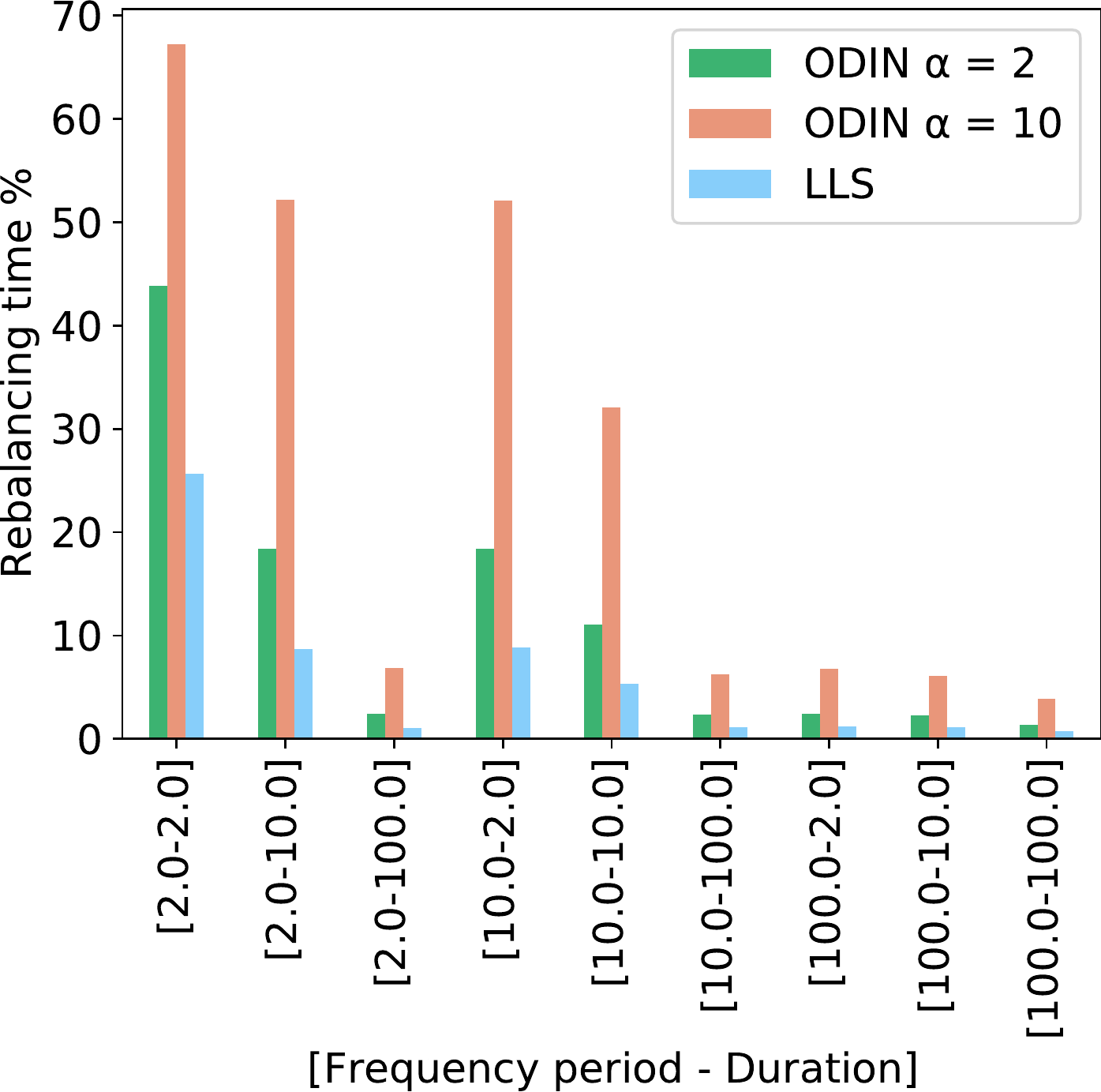}
        \caption{VGG16}
        \label{fig:ov_vgg16}
  \end{subfigure}
%\end{minipage}
\caption{Overhead analysis of ODIN, in comparison to LLS.}
\label{fig:ov}
\end{figure}

\noindent \textbf{Exploration overhead:} Upon detection of interference, both ODIN and LLS begin the rebalancing phase, during which queries as processed serially, until a new configuration of the pipeline stages is found. On average, the number of queries that will be processed serially during a rebalancing phase is 1 for LLS, and 4 and 12 for ODIN with $\alpha=2$ and $\alpha=10$ respectively. Figure~\ref{fig:ov} shows the percentage of time required to rebalance the pipeline stages, for the window of 4000 queries. It is evident that, if the type of interference changes frequently and is short-lived, the overhead of ODIN is higher, as the system is almost continuously in a rebalancing phase. However, when the duration of interference is longer, as the effect of interference on the inference pipeline may be the same, rebalancing may not be triggered, as the selected configuration is already optimal, therefore the rebalancing overhead decreases. Longer frequencies and durations of interference are favored by both ODIN and LLS. \looseness=-1

\subsection{Maintaining QoS with ODIN}
\begin{figure}[!tb]
\centering
\begin{minipage}[]{.48\textwidth}
    \centering
    \begin{subfigure}[]{.97\textwidth}
        \includegraphics[width=0.95\linewidth]{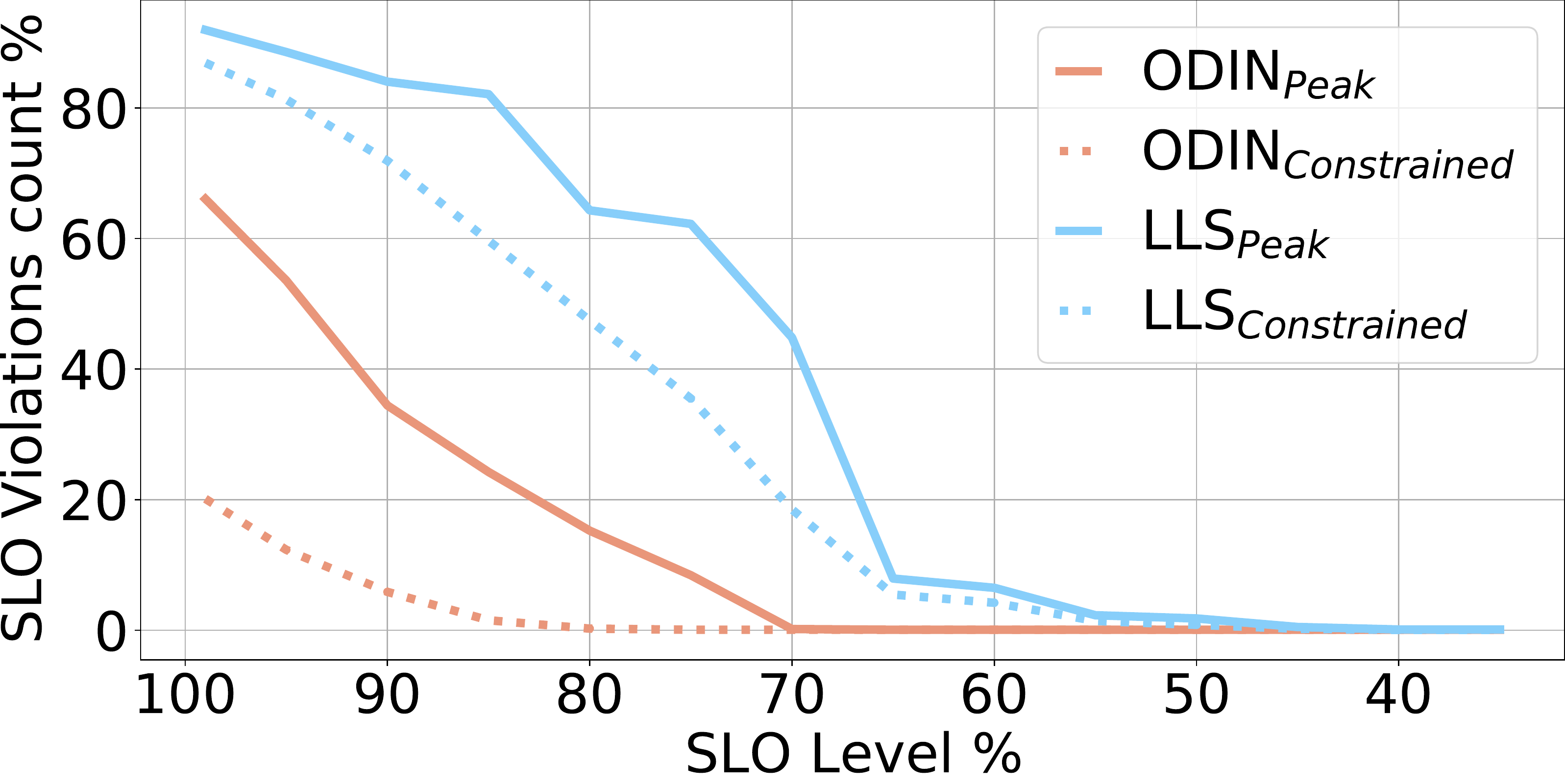}
        \caption{ResNet50}
        \label{fig:lat_scal}
    \end{subfigure}
\end{minipage}
\hfill
\begin{minipage}[]{.48\linewidth}
  \centering
  \begin{subfigure}[]{.97\textwidth}
        \includegraphics[width=0.95\linewidth]{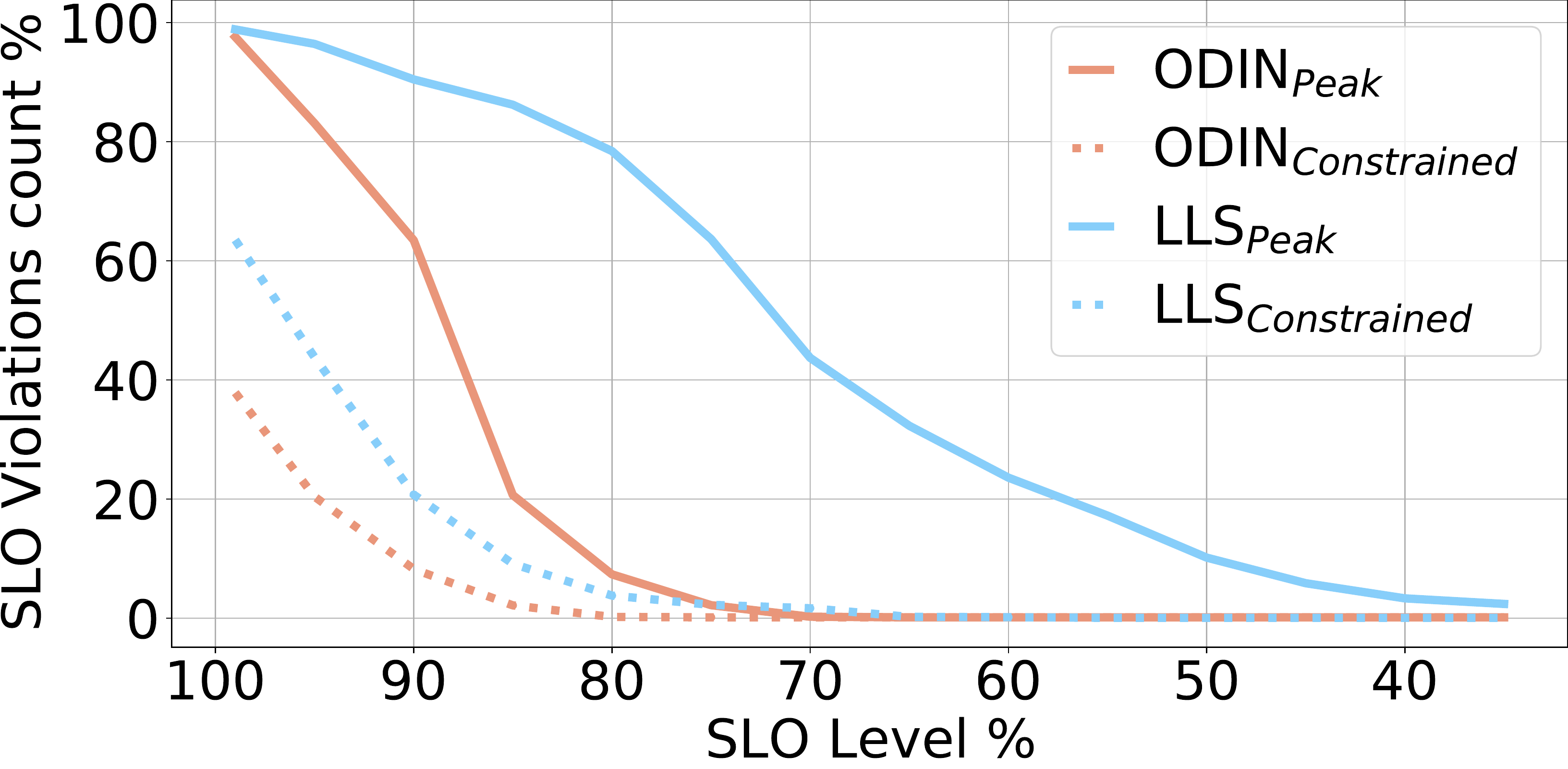}
        \caption{VGG16}
        \label{fig:tp_scal}
  \end{subfigure}
\end{minipage}
\caption{Quality-of-service of ODIN, in comparison to LLS, for different SLO levels.}
\label{fig:slo}
\end{figure}

To evaluate the ability of ODIN to mitigate interference on an inference pipeline, we consider its quality-of-service (QoS) in terms of SLO violations \cite{romero2021infaas, alves2020interference}. We use throughput as the target QoS metric, and consider the SLO level as the percentage of the peak throughput, i.e. the throughput of the inference pipeline when executing alone. We then profile the number of queries which violate this SLO using ODIN and LLS. We additionally compare the SLO violations with respect to the resource-constrained throughput, i.e. the throughput achieved when a colocated workload causes interference, and an optimal configuration of the pipeline is found through exhaustive search. We present the results in Figure~\ref{fig:slo}. Although neither ODIN or LLS are able to offer any performance guarantees, resulting in many violations when the SLO level is strict, ODIN results in less than 20\% of SLO violations for SLO levels lower than 85\%, and can sustain 70\% of the original throughput for any interference scenario, in contrast to LLS, which, in the extreme case of VGG16, violates even an SLO of 35\% of the original throughput. Additionally, the comparison of SLO violations for the SLO set w.r.t. the resource-constrained throughput shows that ODIN is able to find near-optimal configurations in most cases, which are close to those found by the exhaustive search. Our conclusion is that, while ODIN cannot provide any strict guarantee for a set SLO, it can sustain high throughput under looser SLOs and therefore can be an effective solution for overprovisioned systems. For example, an inference-serving system that can tolerate 10\% of SLO violations would require to overprovision resources by 42\% with ODIN, compared to 150\% for LLS. \looseness=-1

% \begin{figure}[!tb]
% \includegraphics[width=\linewidth]{shisha/figs/SLO.png}
% \caption{[Performance delivered by LLS and shisha under different interference scenarios]}\label{fig:slo}
% \end{figure}

\subsection{Scalability analysis of ODIN}
\begin{figure}[!tb]
\centering
    \begin{subfigure}[]{.42\textwidth}
        \includegraphics[width=0.95\linewidth]{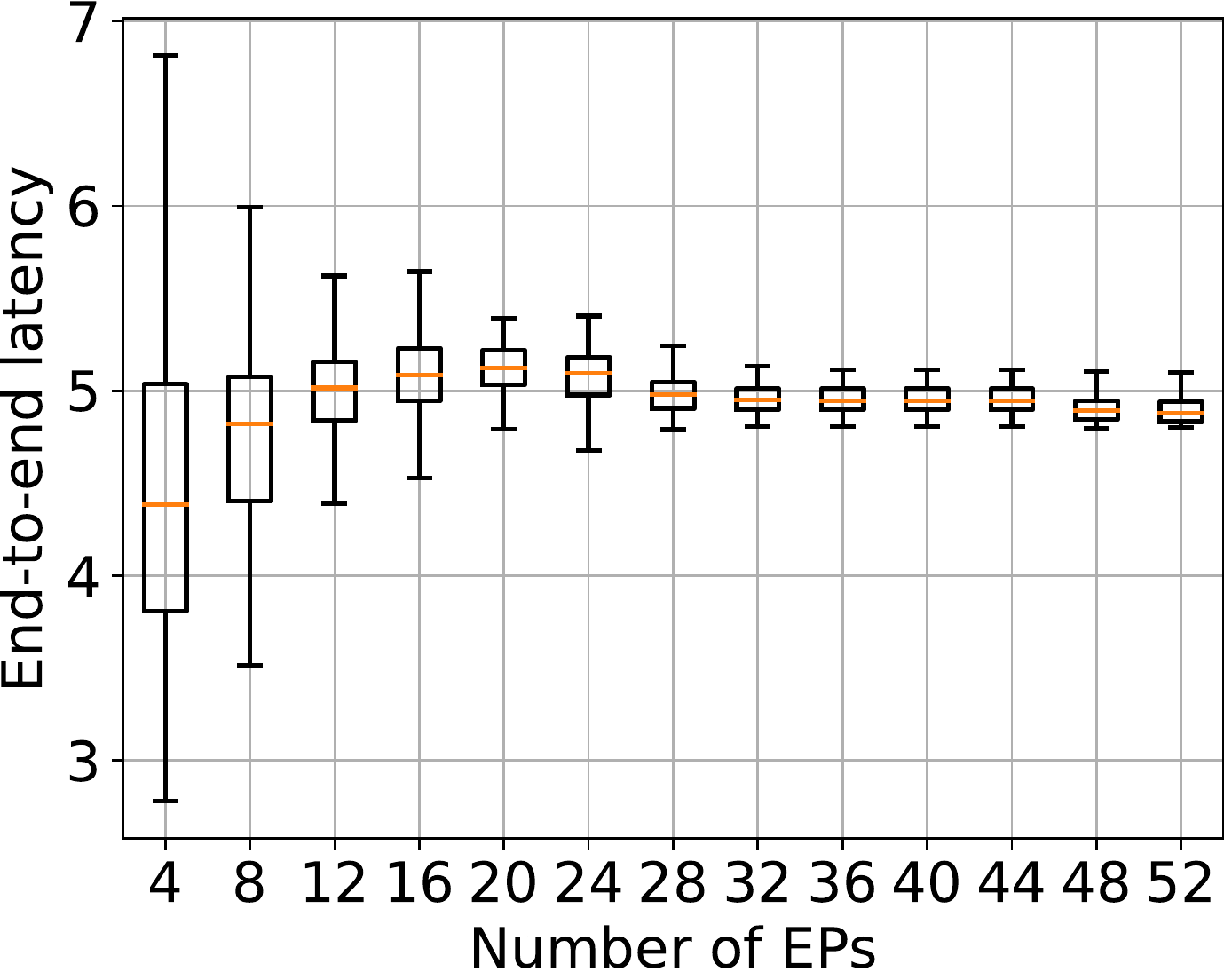}
        \caption{Latency}
        \label{fig:lat_scal}
    \end{subfigure}
\hfill
  \begin{subfigure}[]{.42\textwidth}
        \includegraphics[width=0.95\linewidth]{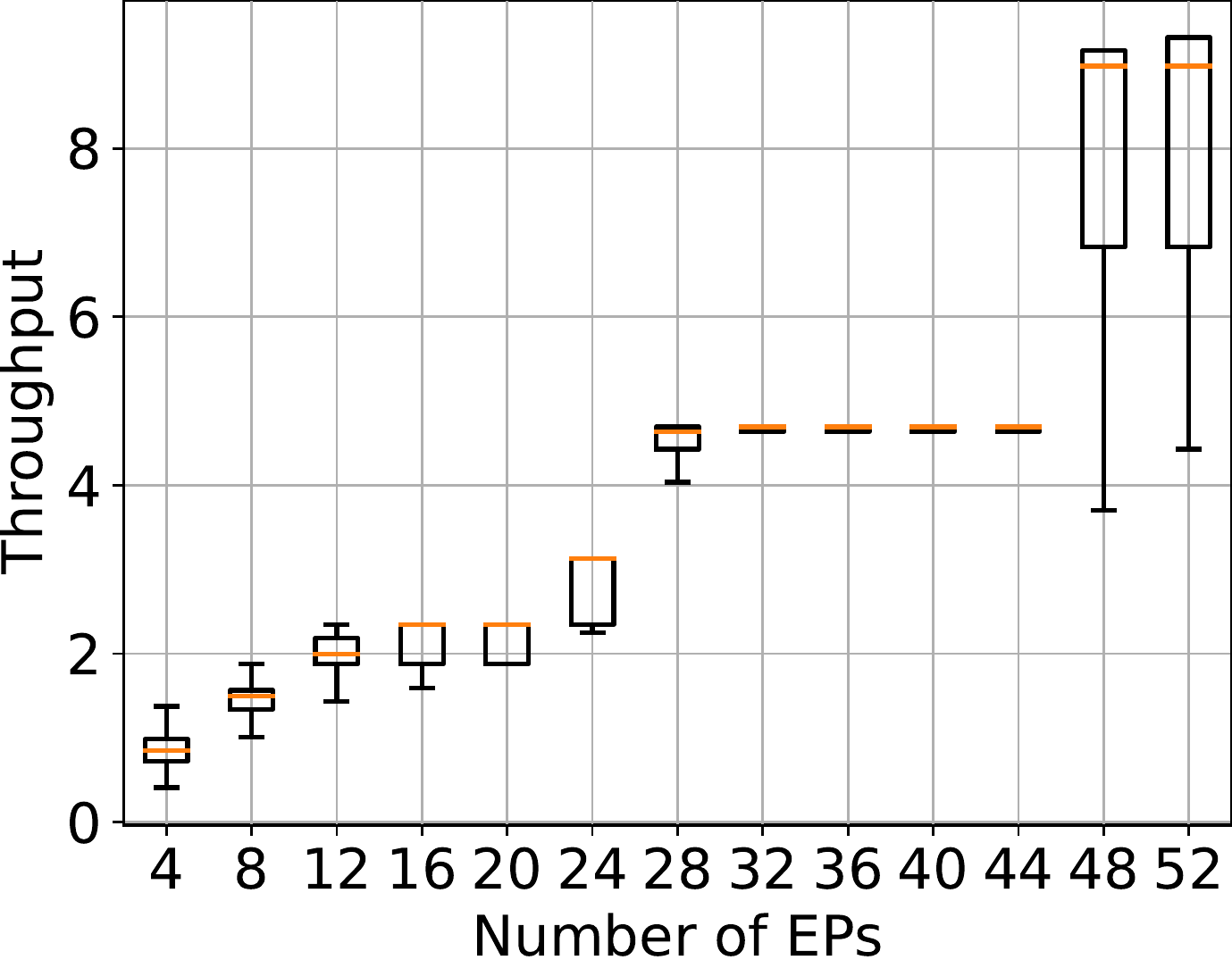}
        \caption{Throughput}
        \label{fig:tp_scal}
  \end{subfigure}

\caption{Scalability analysis of ODIN with ResNet152.}
\label{fig:scal}

\end{figure}

We finally analyze the scalability of ODIN on high numbers of execution places, with deep network models that can run with multiple pipeline stages. For this, we use ResNet152, which consists of 152 layers. We consider, however, residual blocks as a single unit, so the maximum number of pipeline stages ResNet152 could run with is 52. We scale the number of execution places from 4 up to 52, and consider a window of 4000 queries, with interference of a frequency period of 10 and duration of 10 queries. Figure~\ref{fig:scal} shows the latency and throughput of ODIN for the different numbers of EPs. The latency is not affected as the number of EPs increases, therefore ODIN is effective at finding optimal pipeline configurations on multiple execution places. Equivalently, throughput increases with the number of EPs, suggesting high parallelism of the pipeline, and for 52 EPs, the achieved throughput is comparable to the peak throughput of the inference pipeline, under no interference. 

%% file: shisha/conclusion.tex
\section{Conclusion}
In this work, we have proposed ODIN, an online pipeline rebalancing technique that mitigates the effect of interference on inference pipelines. ODIN utilizes the execution times of the pipeline stages to readjust the assignment of layers to pipeline stages, according to the available resources, rebalancing the pipeline. We show that ODIN outperforms the baseline LLS in latency and throughput under different interference scenarios. Additionally, ODIN maintains more than 70\% of the peak throughput of the pipeline under interference, and achieves very low SLO violations compared to LLS. Finally, ODIN scales well with deeper networks and large platforms. ODIN is online and dynamic, and requires minimal information from the inference pipeline, therefore applies to any type of inference pipeline and interference scenario. The abstraction of the hardware into execution places allows ODIN to be applied to different types of hardware platforms. As future work, we plan to parallelize the pipeline during rebalancing, and validate the utility of ODIN on heterogeneous platforms. \looseness=-1

\section*{Acknowledgement}
\noindent This work has received funding from the project PRIDE from the Swedish Foundation for Strategic Research with reference number CHI19-0048.
The computations were enabled by resources provided by the Swedish National Infrastructure for Computing (SNIC) at NSC, partially funded by the Swedish Research Council through grant agreement no. 2018-05973. \looseness=-1